\documentclass[12pt, a4paper]{article} 

\usepackage[utf8]{inputenc} 

\usepackage{graphicx}
\usepackage{epstopdf}
\usepackage{amsthm}

\newtheorem{my_def}{Definition}[section]
\newtheorem{my_th}{Theorem}[section]
\newtheorem{my_lem}{Lemma}[section]
\newtheorem{my_cor}{Corollary}[section]

\usepackage[margin=2.5cm]{geometry}
\usepackage{graphicx} 

\usepackage{booktabs} 
\usepackage{array} 
\usepackage{paralist} 
\usepackage{verbatim} 
\usepackage{subfig} 
\usepackage{amssymb}
\usepackage{amsmath}
\usepackage{lscape} 	
\usepackage{algorithm}	
\usepackage{algorithmic}	
\usepackage{float}

\usepackage{fancyhdr} 
\usepackage[nottoc,notlof,notlot]{tocbibind} 
\usepackage[titles,subfigure]{tocloft} 

\begin{document}

\nocite{*}

\title{Stochastic Price Dynamics Implied By the Limit Order Book}

\author{
Alex Langnau \\ Allianz Investment Management
\and
Yanko Punchev \\LMU Munich
}
\date{} 	

\maketitle

\begin{abstract}
In this paper we present a novel approach to the determination of fat tails in financial data 
by studying the information contained in the limit order book. In an order-driven market buyers and 
sellers may submit limit orders, which are executed if the price touches a pre-specified lower, respectively higher,
limit-price. We show that, in equilibrium, the collection of all such orders - the limit 
order book - implies a volatility smile, similar to observations from option pricing in the 
Black-Scholes model. We also show how a jump-diffusion process can be explicitly inferred to 
account for the volatility smile.
\end{abstract}

\section{Introduction}
The organization of a marketplace where buyers and sellers meet to exchange a well defined asset naturally lies at
the heart of the price discovery process. Traditionally this is done by specialist market makers with a mandate to
match bid and ask quotes from market participants. Such market-order driven ways of trading define the most
``impatient'' form of interactions in the market place - orders are executed immediately as soon as a counterparty
is identified that matches the order even at the expense of getting ``filled'' at a price that is suboptimal to the
initiator of the trade. Such slippage is symptomatic of market-orders and can be viewed as the ``price to pay''
for the impatience to execute. The latter can be avoided by trading \textit{limit-orders} instead of market-orders.
Each limit order includes a price level and a quantity. A seller would specify a pre-defined execution price level
that is typically set \textit{above} the current market price, whereas a buyer would like to purchase 
\textit{below} the current market price at a limit-price of her choice. The important difference to market-orders is
that limit-orders never get filled suboptimally, but may rather not get filled at all or filled only partially in some cases.
Hence the market participant needs to exercise ``patience'' to see an order completed and to be rewarded
with a premium to current levels. The higher the limit-order price level, the higher the potential rewards and the
higher the patience required to see a trade completed in time.

All limit orders are typically collected by the exchange in a limit order book (LOB) that can be accessed by all
market participants. According to P. Jain \cite{Jain}, currently more than half of the world's markets are 
order-driven.

This raises a sequence of interesting theoretical questions about LOBs such as how one should optimally position
oneself for trading in a LOB? What is the information contained in a LOB and when is it in equilibrium?
The recent public availability of LOB data has sparked extensive studies on the structure of LOBs.
The work of J. Bouchaud et al. \cite{Bouchaud} provides useful insights into the shape of the LOB from a statistical
standpoint. The study conducted by A. Ranaldo \cite{Ranaldo} sheds light on aspects of the behavior of market
participants as represented by the change of their impatience preferences as a function of the bid-ask spread.
Further, Z. Eisler et. al in \cite{Eisler} give a detailed look at how the LOB behaves
on different time scales. In \cite{Cont_Talreja} a stochastic model for the dynamics of the LOB in continuous time
has been developed, allowing for simple calibration and explicit calculation of certain probabilities of interest.
M. Bartolozzi \cite{Bartolozzi} proposes a multi-agent model for the dynamics of the LOB, with a particular focus on
capturing key features of high-frequency trading. M. Avellaneda et al. \cite{Stoikov_Avellaneda} also propose a
probabilistic framework for a utility optimizing agent in the context of high-frequency markets. A study conducted
in Spanish equity markets by R. Pascual et. al \cite{Pascual} focuses on what pieces of information of the LOB is
significant and finds that the information is concentrating around the best
bid and ask orders, while orders further out in the LOB do not significantly contribute to the price formation
process. Somewhat similarly, I. Rosu \cite{Rosu2} proposes a dynamic model for order-driven markets with asymmetric
information. He argues that the price impact of market orders is more significant than the impact of limit orders by
an order of magnitude. Furthermore, R. Cont et. al \cite{Cont_Kukanov} study the price impact of order book events
and find a surprisingly simple linear dependence between price changes and an indicator they introduce, which
measures the imbalances between the order flow on the buy and sell sides of the LOB. In \cite{Toth} a study of the
dynamics of different indicators, such as the bid-ask imbalance, is conducted before and after large LOB events and
finds significant dependencies.

The objective of this paper is to study consequences in situations when a LOB is in equilibrium. Following arguments
by Rosu \cite{Rosu}, equilibrium occurs when, at any time, there exists an impatience rate, independent of 
(limit-order price) level, which ``discounts'' limit-sell orders at higher prices in favor of smaller once. Thus an
impatience rate strikes a consistent compromise between higher prices on the one hand, and longer expected 
passage-times to fill on the other hand, throughout the LOB. We study the LOB data of the DAX future and show that
the assumption of a geometric Brownian motion for the price dynamics implies a volatility smile which is reminiscent
of the volatility smile observed in option markets. This allows us to conjecture and re-engineer non-trivial
dynamics underlying the LOB. We show that the assumption of a double-exponential jump dynamics provides a
satisfactory description of the data. Hence this work provides further new evidence for the occurrence of fat tails
in financial data. In addition it provides a novel approach for the determination of jump parameters in finance.

The paper is outlined as follows - first we formalise the general notion of an \textit{impatience rate} - the
quantifier of the trade-off between waiting longer but executing the order at a better price. We then test the 
assertion, that the impatience rate is level-independent, by assuming that the prices follow a geometric Brownian
motion (GBM). We find that the GBM cannot account for a consistent impatience rate, and observe a volatility smile -
if the impatience rate were to be level-independent then limit-orders at higher levels imply an 
ever increasing volatility. Considering the evidence we augment the price process by adding jumps. 
We conclude by providing evidence, that the double-exponential jump-diffusion (DEJD) price process admits an
impatience rate independent of the limit-order level.

\section{Impatience Rate}
In this work we aim to extract empirically testable pieces of information from the LOB, which 
are also pertinent to order-driven markets, and are as much as possible independent of the
model at hand. In reviewing the literature on LOB modelling, a distinction 
between patient and impatient agents has turned out to be the common denominator of 
a much of the research. Specifically in his seminal paper on LOB dynamics, Rosu \cite{Rosu} 
relies explicitly on a parameter $ r $, the impatience rate, which acts as a discounting or 
penalizing factor 
for waiting longer for a better fill price. Rosu further proposes an equilibrium 
model, in which agents maximize their utility according to the following utility function:
\begin{equation}
\label{eq:rosu_utility}
f_t := \mathbb{E}_t[P_{\tau}- r(\tau - t)], \text{ and }
-g_t := \mathbb{E}_t[-P_{\tau}- r(\tau - t)]
\end{equation}
where $ f_t $ is the expected seller utility at time $ t $, $ \tau $ is the time of the limit
order execution, $ P_{\tau} $ is the limit order price
and $ r $ is the common impatience rate of sellers and buyers. Similarly $ -g_t $ is the buyer's utility. 
Regardless of specific price dynamics assumptions and LOB model, a limit order far away from the 
current price will take longer to fill, so $ r $
weighs the benefit of a better fill price, which increases utility, by penalizing for waiting longer and
decreasing utility. So the impatience rate should
quantify this fundamental trade-off between waiting longer to execute at a better price and 
waiting less, but executing at a worse price. Consequently, if the LOB is in a state of equilibrium, and if 
capital markets are efficient, the impatience rate should be the same across limit-order levels.

I. Rosu in \cite{Rosu} shows that such an equilibrium exists in his theoretical framework, and empirical testing
should provide evidence whether the LOB is efficient, or at least imply what the price process
should look like, if the assumption of information efficiency were to hold.
Despite playing such a central role in many works on limit-order markets, the properties of the impatience 
rate $ r $ have not been examined.

The utility function as given in \eqref{eq:rosu_utility} is somewhat unfortunate. On the 
on hand, it depends on the absolute level of the limit order price, $ P_{\tau} $, which leads 
to different results even in Rosu's model for markets, which are equivalent up to a price 
scaling constant. Also, this approach suppresses an important characterization of the limit 
order, namely its distance to the current best offer, or at least to the mid price. 
In \eqref{eq:rosu_utility} this is expressed only indirectly through the expected 
hitting time $ \mathbb{E}[\tau] $. At this point another drawback of this particular utility 
becomes apparent - the fact that $ \mathbb{E}[\tau] = \infty $, if the asset price is 
modelled by an exponential Brownian motion without drift, or if the drift is 
in a direction away from the submitted limit order.\footnote{For a discussion of how 
the GBM should be specified, so that the expected hitting time is finite
cf. M. Yor et. al \cite{Yor}.}

Since \eqref{eq:rosu_utility} depends on the absolute level of the price, and would generally be 
infinity for a Brownian motion price process we investigate a different utility function. 
We consider the currency value of the \textit{distance-to-fill}, i.e. how much 
the best offer has to travel until it meets a trader's limit order, discounted by the impatience rate
in a quite literal sense:
\begin{equation}
U_D(r)=|D|\mathbb{E}[\exp(-r\tau_{S+D}) | S_0=S]
\label{eq:the_utility}
\end{equation}
$ D\in \mathbb{R} $ is the distance from the limit order to the best offer $ S_0 $ at time $ t=0 $. 
Further $ \tau_{S+D} $ is the first hitting time of the price process $ S_t $ describing the evolution
of the best offer, started in $ S_0 $:
\begin{equation*}
\tau_{S+D}:=\inf_{t\ge 0}\lbrace S_t = S + D \rbrace.
\end{equation*}

\section{A Pure Diffusion Setting}

We first adopt a model in which the asset price $ (S_t)_{t\ge 0} $ is a geometric Brownian motion (GBM).
Specifically let $ \left( \Omega, \mathcal{F}, \left( \mathcal{F}_t \right)_{t\ge 0},
\mathbb{P} \right) $ be a filtered space
\footnote{In the rest of this paper we will always assume, that any stochastic process
we introduce is adapted to this generic filtered space and its dynamics 
are given with respect to the physical probability measure $ \mathbb{P} $.}, 
and $ (S_t)_{t\ge 0} $ be a stochastic process on this filtered space, whose
dynamics are given by the stochastic differential equation (SDE)
\begin{equation*}
\text{d}S_t = S_t \mu \text{d}t + S_t \sigma \text{d}B_t \;, \quad S_0 = S
\end{equation*}
where $ S, \sigma >0 $, $ \mu \in \mathbb{R} $ and $ (B_t)_{t\ge 0} $ is a standard Brownian motion, started
in 0. The SDE has the exact
solution \footnote{Cf. \cite{Shreve}}:
\begin{equation}
S_t = S \exp\left( \left(\mu-\dfrac{\sigma^2}{2}\right)t + \sigma B_t \right)
\end{equation}
We define the asset price log-returns over an interval $ \Delta t >0 $ by
\begin{equation*}
l(\Delta t):=\ln\left(\dfrac{S_{t+\Delta t}}{S_t}\right),
\end{equation*}
and establish the following results:
\begin{align}
\label{eq:gbm_mean}
\mathbb{E}[l(\Delta t)] & = \mathbb{E}\left[\left(\mu - \dfrac{\sigma^2}{2}\right)\Delta t
+ \sigma B_{\Delta t} \right] = \left(\mu - \dfrac{\sigma^2}{2}\right)\Delta t\\
\label{eq:gbm_var}
\mathbb{V}[l(\Delta t)] & = \mathbb{E}\left[\left(\left(\mu - \frac{\sigma^2}{2}\right)\Delta t
+ \sigma B_{\Delta t} \right)^2 \right] - \left(\left(\mu - \frac{\sigma^2}{2}\right)\Delta t\right)^2
= \sigma^2 \Delta t
\end{align}
Further we note that the Laplace transform of the first hitting time at level $ z \in \mathbb{R} $
$$ \hat{\tau}_z :=\inf_{t\ge 0 }\lbrace B_t =z  \rbrace $$
of a standard Brownian motion with drift $ \hat{\mu} $
$$ B_t^{(\hat{\mu})} := \hat{\mu} t + B_t $$ 
is given by\footnote{Cf. \cite{Borodin} (page 223, 2.0.1).}

\begin{equation}\label{eq:th_1}
\mathbb{E}[\exp(-r \hat{\tau}_z)] = \exp\big(\hat{\mu}z - |z|\sqrt{2r+\hat{\mu}^2} \big)
\end{equation}
for some $ r>0 $. 
Finally a simple transform is needed to obtain a closed formula for the expected utility:
\begin{align}
\tau_{S+D} & =\inf_{t\ge 0}\lbrace S_t \ge S + D \rbrace = 
\inf_{t\ge 0}\lbrace S_t = S + D \rbrace \nonumber \\
&=\inf_{t\ge 0}\left\lbrace S \exp\left( \left( \mu - \frac{\sigma^2}{2}\right)t + \sigma B_t \right) =
S+D \right\rbrace \nonumber \\
& = \inf_{t\ge 0}\Bigg\lbrace \underbrace{\frac{\mu-\frac{\sigma^2}{2}}{\sigma}}_{=:\hat{\mu}} t + B_t = 
\underbrace{\frac{\ln\left(\frac{S+D}{S}\right)}{\sigma}}_{=:z}\Bigg\rbrace
 \label{eq:GBM_z} \\
& = \hat{\tau}_z, \quad S\in \mathbb{R}_{+} \quad D \in ]-S,\infty[ \nonumber
\end{align}
Now $ \hat{\tau}_z $ is the first hitting time of a standard Brownian motion with drift $ \hat{\mu} $
for the hitting level $ z\in\mathbb{R} $. Notice
that we allow $ z<0 $, so that we can use the same result for hitting levels above the 
current price, as well as below the current price. In particular we can use the same formula
for the bid and for the ask side of the book. Now we are in a position to compute explicitly
for $ D \in \mathbb{R} \setminus \lbrace 0 \rbrace$, $ r, S >0 $:
\begin{align}
U_D(r) & = |D| \mathbb{E}[\exp(-r \tau_{S+D}) | S_0 = S ] \nonumber \\
& = |D| \mathbb{E}[\exp(-r\hat{\tau}_z)] \nonumber \\
& = |D| \exp\left(\hat{\mu}z - |z|\sqrt{2r+\hat{\mu}^2} \right) \nonumber \\
& = |D| \exp\left(\frac{\left(\mu-\frac{\sigma^2}{2}\right)\ln\left(\frac{S+D}{S}\right)}{\sigma^2}
- \frac{\left\vert\ln\left(\frac{S+D}{S}\right)\right\vert}{\sigma}
\sqrt{2r+\left(\frac{\mu-\frac{\sigma^2}{2}}{2}\right)^2} \right)
\label{eq:gbm_utility}
\end{align}
In order to calibrate our model to market data, we need to estimate the drift and
the volatility of the GBM. We do this by estimating the sample mean and standard deviation of 
the log-returns market mid-prices at equidistant time points, with a constant 
time interval of $ \Delta t $. Recall that $ M_t $, the mid price at time $ t $,
is simply the mid-point of the ask price at level zero $ A_t^{(0)} $ and the bid price
at level zero $ B_t^{(0)} $:
\begin{equation*}
M_t:=\dfrac{B_t^{(0)} + A_t^{(0)}}{2}, \quad B_t^{(0)}, A_t^{(0)} > 0
\end{equation*}
We do a rolling estimate for each data point $ i $ in equidistantly 
spaced LOB (i.e. $ t_i - t_{i-1}= \Delta_t $) using a standard point estimate of the
sample mean, based on the last thirty observations prior to the current point:
\begin{equation*}
\overline{m}_i(\Delta t) :=\dfrac{1}{30}\sum_{k=1}^{30}\log\left(\dfrac{M_{i-k}}{M_{i-k-1}}\right)
\end{equation*}
Similarly we estimate the standard deviation of the log-returns by:
\begin{equation}
\overline{s}_i(\Delta t) := \sqrt{\dfrac{1}{29}\sum_{k=1}^{30}
\left(\log\left(\dfrac{M_{i-k}}{M_{i-k-1}}\right)-\overline{m}_i(\Delta t)\right)^{2}}
\end{equation}
Notice that $ k $ starts at one, ensuring that our current estimate of the sample mean and standard deviation
uses only past data. 
Recalling the expressions for the expected value \eqref{eq:gbm_mean} and the variance \eqref{eq:gbm_var}
of a GBM, we first substitute $\overline{s}_i(\Delta t)^2$ for 
$ \mathbb{V}[l(\Delta t)] $ in \eqref{eq:gbm_var} to obtain the simple rolling estimate:
\begin{equation}
\sigma^2 \Delta t =  \overline{s}_i(\Delta t)^2
\end{equation}
Plugging this expression in \eqref{eq:gbm_mean}, and substituting $ \overline{m}_i(\Delta t) $
for $ \mathbb{E}[l(\Delta t)]$, we estimate:
\begin{align}
& \overline{m}_i(\Delta t) = \left(\mu - \dfrac{\overline{s}_i(\Delta t)^2}{2}\right) \Delta t
\nonumber \\
\Rightarrow &  \mu \Delta t = \overline{m}_i(\Delta t) + \dfrac{\overline{s}_i(\Delta t)^2}{2}
\end{align}
Now our model is thoroughly specified and we are in a position to calculate the expected 
utility at any point in time, given the value of the impatience rate at that point. Since, 
however, the impatience rate is precisely what we would like to estimate, we need 
an additional assumption about the whole setting. Considering that 
the LOB is in an equilibrium when the expected utility of all participants on each side of the 
book is the same, so no one has an incentive to put in their order at a different level\footnote{Cf. 
\cite{Rosu}}, 
we assume that the expected utility is 
constant over short periods of time, and that it evolves fairly smoothly in time, as
the expectations of the market participant on each side of the book for the 
future direction of price movements changes through time. By starting at some reasonable value
for the expected utility we can fit the impatience rate stepping through time, with 
the objective of keeping the utility stepwise constant achieving a smooth fit. 

A clearer specification of the algorithm impatience rate estimation 
by fitting the expected utility to market data in a GBM framework is due. 
The inputs required are the following:
\begin{itemize}
\item[$\bullet$] $ N $ - the number of data points in the reconstructed LOB with equidistant spacing of
$ \Delta t $. The results presented in this paper are for $ \Delta t = 30 $ seconds.
\item[$\bullet$]  ``steps" - indicates the number of data points, over which the expected 
utility is to be held constant, in order to minimize an error function. In this paper we present
results for ``steps"=2, in order to compare the results better to the DEJD model;
\item[$\bullet$] m - a vector of size $ N $, containing the rolling estimates for the 
mean of the observed log-returns where the time interval between observations is 
$ \Delta t $. Note that $ m_j $, the estimate at time $ t_j $, is based on the log-returns 
at thirty observations prior to $ t_j $, so no peeking in the future is allowed;
\item[$\bullet$] s - a vector of size $ N $, containing the rolling estimates for the 
standard deviation of the observed log-returns where the time interval between observations is 
$ \Delta t $, where again estimates are based on a past data only;
\item[$\bullet$] $ U_0 $ - an initial estimate for the expected utility;
\item[$\bullet$] $ r_0 $ - an initial estimate for the impatience rate;
\end{itemize}
The outputs are two vectors $ U $ and $ r $, each of size $ N $, containing 
the expected utility and the impatience rate through time. 

\begin{algorithm}
\caption{Estimation of the impatience rate in a GBM setting}
\label{GBM_algo}
\textbf{Inputs:} N, steps, m, s, $ U_0 $, $ r_0 $
\begin{algorithmic}[1]
\STATE $ r $ = $ r_0 $
\FOR {i=1:steps:N } 
	\WHILE{$ \text{error(r)}>\epsilon \; \& \; \text{Tolerance} > \delta $}
		\STATE $ r_i := \min_{r>.01}\lbrace \text{error(r)} \rbrace $ 
		\STATE \textbf{WHERE} error(r) is 
		\FOR {j=i:steps+1} 
			\STATE	calculate $ U_j $ according to \eqref{eq:gbm_utility} using $ m_j $, $ s_j $, $ r $
		\ENDFOR
		\STATE error(r) = $ \sum_{j=i}^{\text{steps}+1} 
							\left(\dfrac{U_j - U_{j-1}}{0.5(U_j + U_{j-1})}\right)^2 $	
	\ENDWHILE
\ENDFOR
\end{algorithmic}
\textbf{Outputs:} $ U $, $ r $
\end{algorithm}

The particular choice of the error function in row 9 is a natural one - it collects
the differences in utilities step by step, scaling each error by the mean of both adjacent
utilities, ensuring the minimization algorithm would not just choose a large $ r $ to converge
by essentially driving all utility down to zero. Particular care has to be taken 
when choosing $ \epsilon $, $ \delta $ and other minimization algorithm break 
criteria in order to ensure timely convergence. Also one has to consider what $ \Delta t $
to choose in order to reduce the number of time steps of the immense data set the LOB
offers per day (in our case about 450,000 observations per day in the LOB organization of 
Figure \ref{fig:formated_LOB} in Appendix C). Further when choosing ``steps", one faces 
the trade-off between smoothness of $ U $ for an increasing number of ``steps" on the one hand,
against potential convergence problems, as well as a more noisy $ r $.
\section{Results For a Pure Diffusion}
We find, that in a GBM setting, the impatience rate $ r $ is not constant across levels, meaning that either
the market is not in equilibrium, as contended in \cite{Rosu} and dictated by the efficient markets
hypothesis, or that the GBM framework is an inadequate cannot describe the LOB equilibrium. We
further observe a volatility smile (see Figure \ref{fig:GBM_smile}), which leads us to assume a jump-diffusion
process for the asset price dynamics. 

All results presented here are based on the following parameters: 
$ \Delta t $ = 30 sec; ``steps" = 2; $ r_0=1 $ and $ U_0 $ is calculated 
according to \eqref{eq:gbm_utility} for $ r= r_0=1 $. It should be noted, that we achieved very similar 
results for ``steps=15" and for ``steps=30", indicating the fitting algorithm is 
independent of the particular time-stepping procedure. The results are also the same for different
starting $ r_0 $ and $ U_0 $, providing evidence that the problem is well-defined and fitting procedure is also 
well-posed.

Our results are based on nearly three months of LOB data of the DAX future 
from June to August 2010. Analysis of the data revealed properties of the impatience
rate and the expected utility which have been very consistent throughout trading days.
We will illustrate them based on a representative day of our data set - 20 August, 2010 - 
and refer the reader to the appendix for descriptive statistics for each day.

\begin{figure}[!ht]\centering
\includegraphics[scale=0.9]{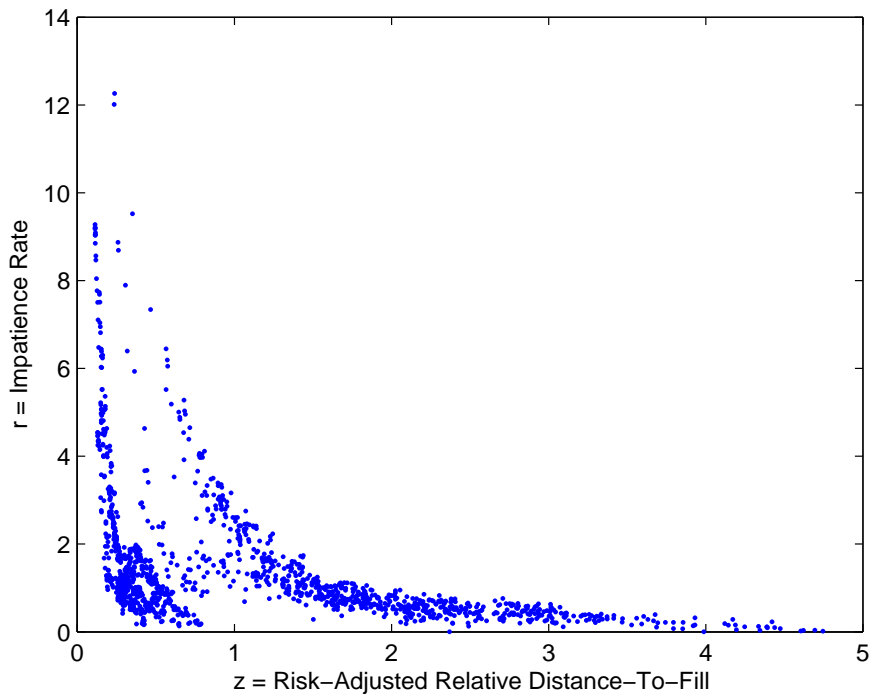}
\caption{Impatience rate ($ r $, y-axis) against risk-adjusted relative distance-to-fill
($ z $, x-axis) for the sell side of the book. The fitting procedure reveals 
a power law for $ r $, 
which is driven by both - the estimated volatility $ s $ and the absolute distance-to-fill
$ D $. (FDAX, August 20, 2010)}
\label{fig:gbm_imp_rate_z}
\end{figure}

In Figure \ref{fig:gbm_imp_rate_z} we present the impatience rate $ r $ as a function of the 
risk-adjusted relative distance-to-fill 
\begin{equation}
z=\dfrac{\log\left(\dfrac{S+D}{S}\right)}{\sigma},
\end{equation}
for the sell side of the LOB. If 
$ r $ were to be constant, strains of flat lines, each strain indicating a different market mode,
should be observable. 
Instead, for different levels of expected utility, the impatience rate assumes a power law of the form 
\begin{equation}
r=\dfrac{1}{z^{\alpha}}
\end{equation}
for some $ \alpha>0 $.

\begin{figure}[!ht]\centering
\includegraphics[scale=0.9]{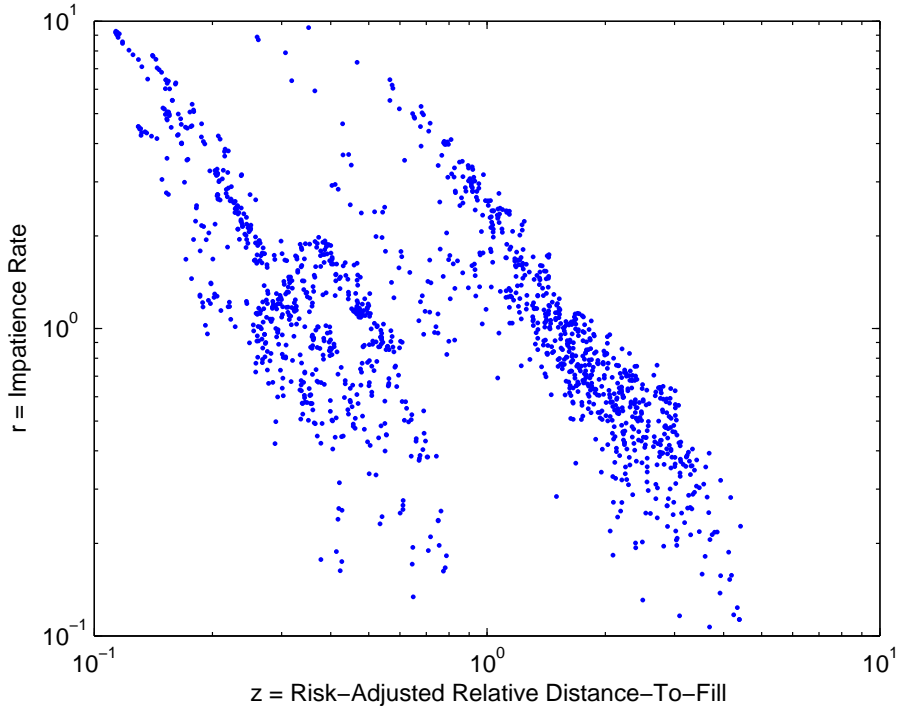}
\caption{$ \log_{10} $ of the impatience rate ($ r $, y-axis) against 
$ \log_{10} $ of the risk-adjusted relative distance-to-fill
($ z $, x-axis) for the sell side of the book. Notice the different linear
(exponential on a linear scale) strains, which indicate the same 
power relationship between $ r $ and $ z $ for different levels of the expected
utility. (FDAX, August 20, 2010)}
\label{fig:GBM_z_r_loglog}
\end{figure}

The relationship demonstrated in Figure \ref{fig:gbm_imp_rate_z} provides evidence, that the impatience rate is
primarily a function of 
the estimated volatility and the distance-to-fill $ D $. A first conclusion is that in the 
GBM model the impatience rate is not constant across different levels. In particular 
it cannot extract solely the information content about the trade-off between a better fill price and a longer
expected waiting time, but much rather it still contains information about the volatility and 
the distance-to-fill.
Next, in Figure \ref{fig:GBM_z_r_loglog} we show a log-log plot which suppresses the power law relationship between 
$ r $ and $ z $ and reveals the different strains of this relationship for different levels of expected utility.
The relative level of expected can be interpreted as indicative of prevailing market sentiment and therefore
different levels of expected utility represent different market modes, and sharp changes in the absolute value
of utility point to a sentiment shift.

\begin{figure}[!ht]\centering
\includegraphics[scale=0.9]{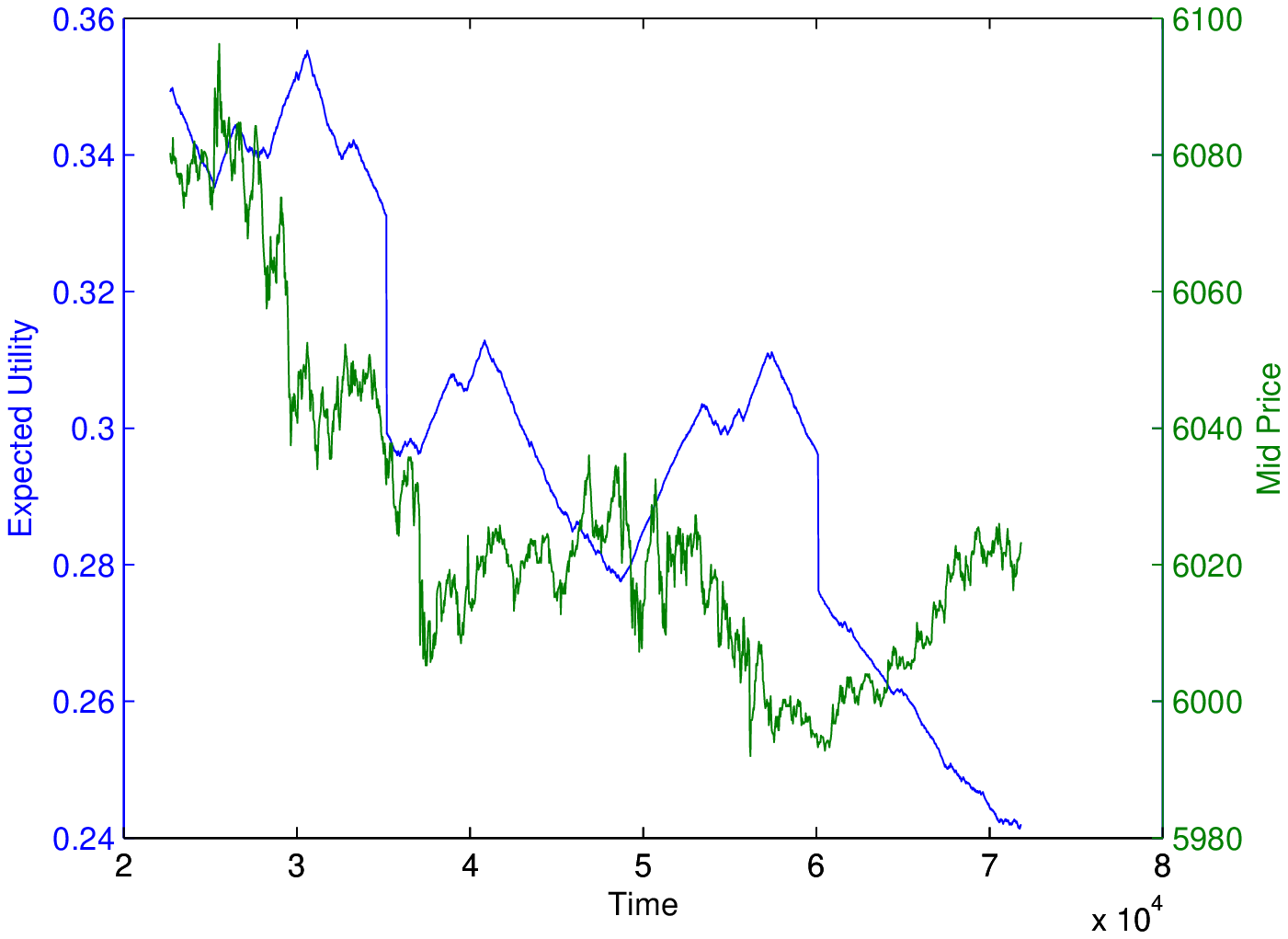}
\caption{The expected seller utility $ U $ (left axis) and the mid-price 
(right axis) through time (x-axis) on a single trading day. Notice the initial drop
from the initial estimate to a level which is reflecting the actual expected utility
given the impatience rate $ r $, and the subsequent smoothness of the evolution 
of the utility. Jumps indicate a change in the utility level and correspond 
to a different linear strain in the relationship between $ r $ and $ z $, 
as demonstrated in the previous figure. (FDAX, August 20, 2010)}
\label{fig:GBM_u_mid}
\end{figure}

In figure \ref{fig:GBM_u_mid} we present the expected utility compared to the mid-price through time. 
Changes in the direction of the expected utility, indicate numerical
instability and can be interpreted as an indication, that a shift in sentiment of the respective
market participants (here - the sellers), is taking place. A more detailed analysis showed
that while indeed price direction and expected utility changes are concurrent, the latter 
is not a reliable predictor of the former.

\begin{figure}[!ht]\centering
\includegraphics[scale=0.9]{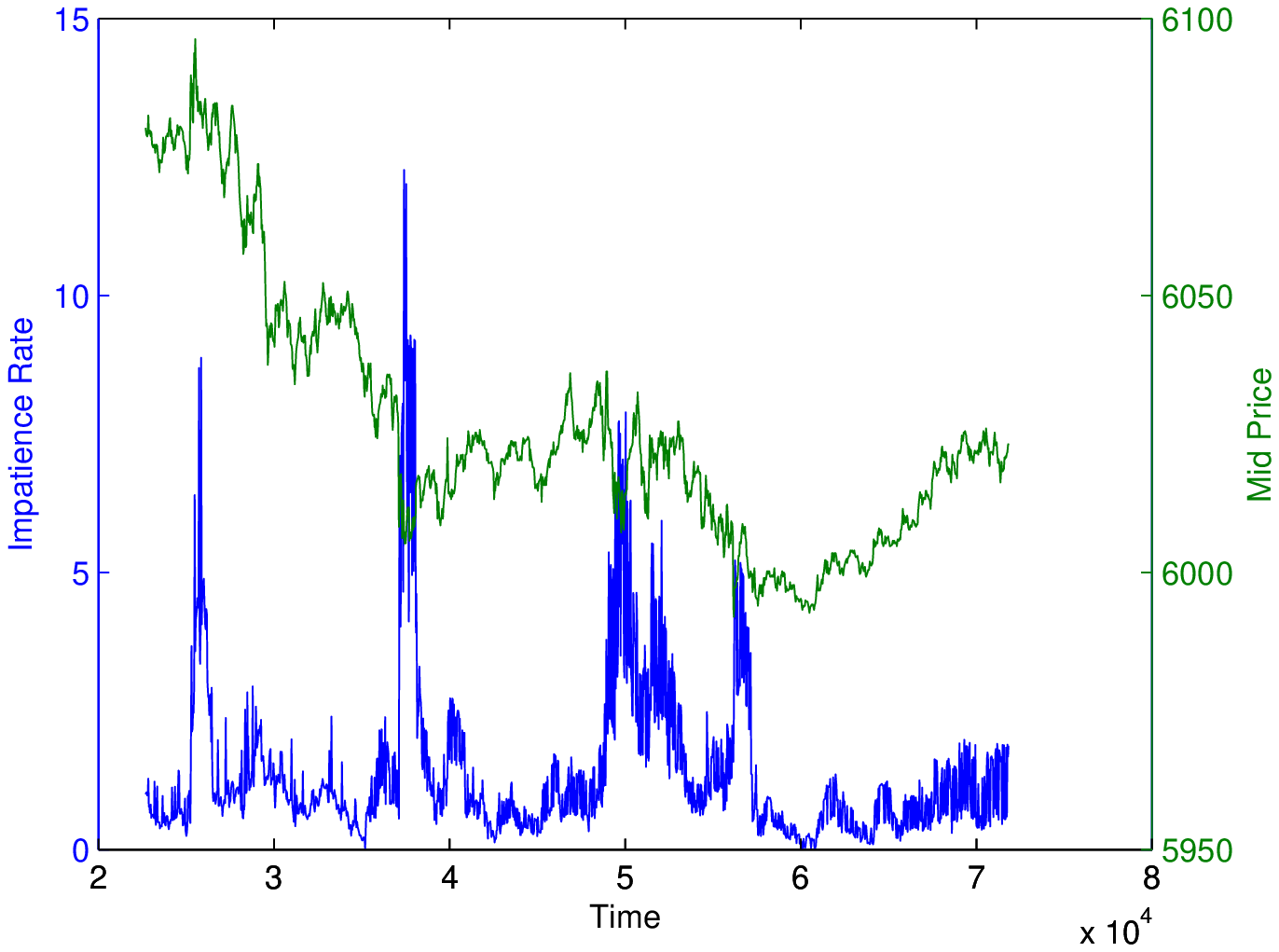}
\caption{The seller impatience rate $ r $ (left axis) and the mid-price 
(right axis) through time (x-axis) on a single trading day. Intuitively a rise in 
the seller impatience rate should precede or accompany a fall in the price. 
(FDAX, August 20, 2010)}
\label{fig:GBM_r_mid}
\end{figure}

In Figure \ref{fig:GBM_r_mid} a comparison of the impatience rate and the mid-price is given.
One of the objectives of this research is to conclude whether the impatience rate is somehow
indicative of future price movement, or if it contains any other piece of useful information.
The intuition is that 
on, per example, the sell side of the book, an increasing impatience rate $ r $ indicates, that
sellers are becoming eager to get out of their holdings, so a price deterioration can be
anticipated.

While testing has revealed, that sharp changes in the mid-price are accompanied 
a shift in the opposite direction of the sell-side impatience rate, the direction of 
$ r $ is not a reliable indicator of future price movements, since there are false outbreaks. Also
price changes, which have not occurred abruptly enough, may not be captured by a significant 
change in $ r $ and thus missed. Notice also how the impatience tends to fluctuate, around
a stable state, when the price is trending sideways. This is a particular consequence of the 
fact that the impatience rate is a function of the distance-to-fill - the fluctuations 
indicate that the impatience rate needs to assume a different value for new values of 
$ D $, in order to keep the expected seller utility smooth. As a result regression
analysis has showed no significant relationship between the estimated impatience rate 
and future mid-price returns.
The results for the buy-side of the LOB are similar and a side-by-side comparison of sell and buy side 
results can be seen in the appendix.

\section{Volatility Smile}
\begin{figure}[!ht]\centering
\includegraphics[scale=0.9]{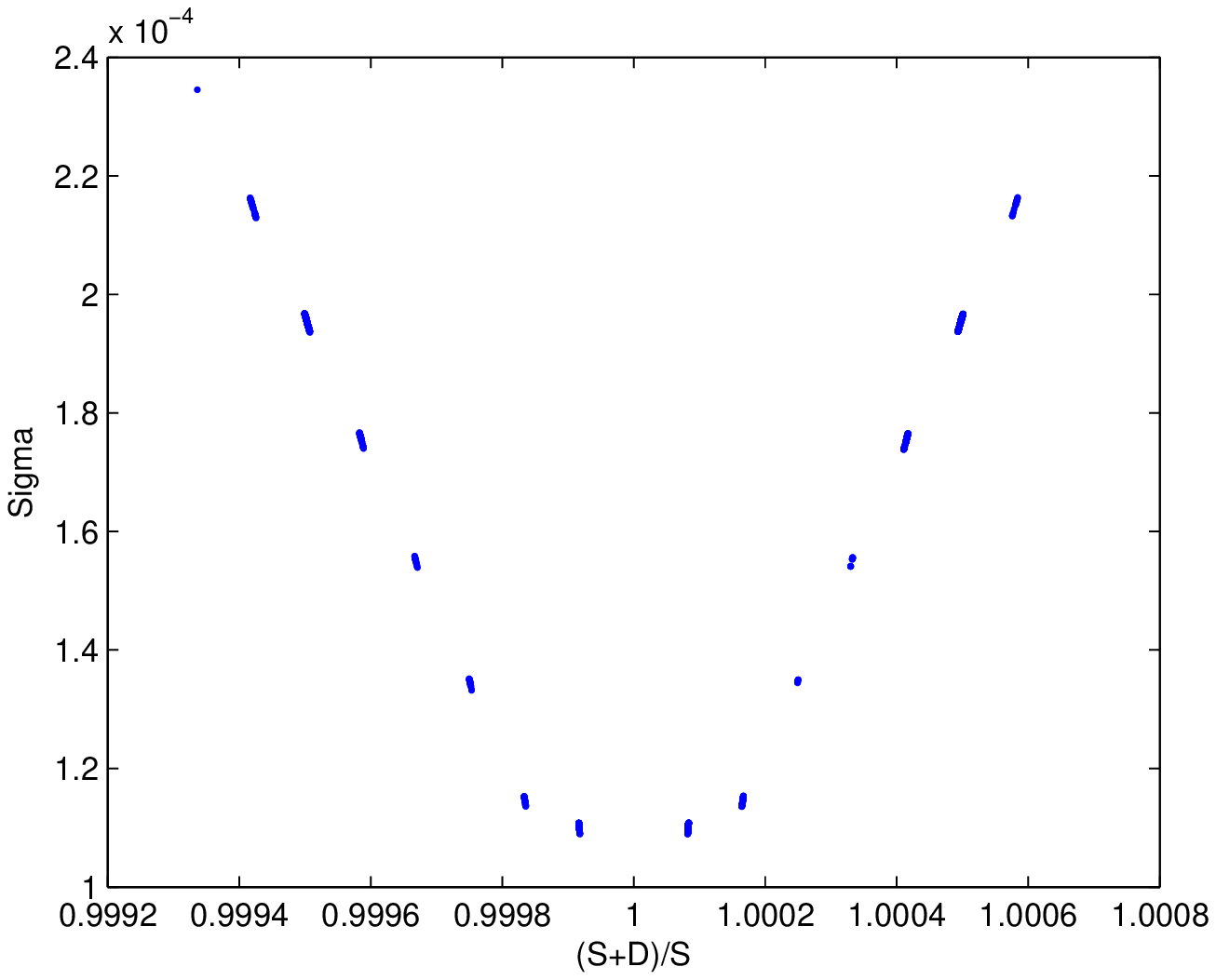}
\caption{The volatility smile, as implied by the LOB. A scatter plot of the implied volatility
$ \sigma $ (y-axis) and the relative distance-to-fill $ (S+D)/S $ (x-axis)
form what resembles a volatility smile as known from equity options pricing. The plot
reveals a whole day's worth of data. $ \sigma $ values to the left of one are 
derived from  the bid-side of the LOB, and to the right - from the ask-side. This instance of the smile was 
derived for $ U=0.2355 $ and $ r=0.5 $
(FDAX, August 20, 2010)}
\label{fig:GBM_smile}
\end{figure}

The results reveal that the impatience rate is not constant across LOB levels and 
in this model framework cannot be used as a consistent quantifier of the 
trade-off between a better fill price a longer expected time-to-execution. Further analysis
suggested that the impatience rate in a GBM model is a function of the distance-to-fill
and of the estimated volatility. Intuitively the market
participants, who place their orders further out in the LOB, imply a much higher
volatility than the observed. It seems as if traders who place orders at higher levels in the 
LOB are betting on a sharp price change in the desired direction. 
In order to analyse this further, we take a look at the \textit{LOB implied volatility},
which we define to be the volatility $ \sigma $ for which 
\begin{equation}
U_{D}(r)=U_{D}(r, \sigma)= c
\end{equation}
for some positive constant $ c $. As we have already established, the impatience rate
depends on $ D $, and on $ \sigma $, so we would like to separate and measure the 
two effects. To this end we choose an arbitrary, but reasonable, constant $ c $ for 
the utility level and also fix the impatience $ r $ at some constant level. Both are held constant throughout
a whole trading day, and we fit for $ \sigma $, to derive the implied volatility which 
keeps the impatience rate constant at $ r $ and the expected utility level $ U_D(r,\sigma) $
constant at a level $ c $ for the entire trading day. What we discover is 
a \textbf{volatility smile}, being implied by the LOB, as is demonstrated in Figure
\ref{fig:GBM_smile}. Essential this shows, that the GBM underweights limit orders at higher levels -
the process' volatility is insufficient for it to reach these limits in due time, so the volatility has to be tuned
up to account for the dynamic equilibrium, and for market efficiency to hold.

The empirical work conclusively shows, that in a GBM setting, there is no 
unifying impatience rate across the different levels in the LOB. It has turned
out that it is a function of both - the volatility term $ \sigma $ of the 
stochastic process, and of $ |D| $ - the distance-to-fill. It is inversely related 
to the volatility, i.e. the higher the volatility, the lower the impatience rate, 
on either side of the book. It means that it becomes ``easier'' for the process to reach
an order, which is at a higher level in the book, which is in turn necessary due 
the assumption that the expected utility is the same at all levels 
of the book, and evolves smoothly through time. 

In the same way, the impatience rate is a function of the distance-to-fill. It plays
the role of a ``volatility-compensator'', keeping the expected utility $ U $ 
constant across levels for different values of $ |D| $ . The impatience rate needs
to be very small for large $ |D| $ in order to compensate for the decrease
in utility, as $ \tau $ becomes too big. Since the volatility is fixed by 
empirical observations, and $ \mathbb{E}[\tau] $ is known\footnote{Cf. \cite{Borodin}.}
 to grow exponentially with 
rising distance, it is a direct consequence of the GBM framework, that $ r $ 
follows an exponential decay law for an increasing $ |D| $ and a falling $ \sigma $.

The inverse power-law between the impatience rate and the observed volatility is 
especially clearly observable in the volatility smile. This surprising discovery
highlights the fact that in order to achieve utility equilibrium one has to 
tune up the process' volatility, so that limit orders placed further away from the
mid-price have a higher chance of execution. Intuitively one may argue, that 
market participants place limit orders further out in the LOB, because they anticipate 
a large block of orders being placed at once, so that lower level limit orders get immediately
filled, leaving the agent's order as the best or even 
filling it. 

A critical argument for jump augmentation is based on the observed ``volatility smile", a phenomenon
well known from option pricing.\footnote{Cf. \cite{Derman}, \cite{Sircar} 
on more about the volatility smile in the context of option pricing. }
There are already many models in options theory which account for the smile. 
Regarding the principle idea of the solution, there are two broad classes of 
approaches to incorporate this phenomenon in the price process. One is to
assume a stochastic process for the volatility term in front of the 
Brownian motion. The other is to add a jump process to the Brownian motion. While 
it might seem somewhat natural that the volatility itself should also be a process 
rather than a constant term, the dynamics that are usually used to model the 
evolution of the volatility through time are not necessarily intuitive.\footnote{Cf.
 \cite{Brockwell} } On the other hand, the idea, that price processes
have jumps is considered characteristic of financial time series\footnote{Cf. \cite{Brockwell},
 \cite{Brockwell_Davis}, \cite{Brockwell_Davis_2},\cite{McNeil} }
and is readily
 observable\footnote{Cf. \cite{Brockwell},\cite{Maekawa}, \cite{Thierry} }. In combination with the 
 intuition, that limit order traders position themselves at the outer levels in
 anticipation of a block-trade, or a sharp price movement, we are lead to extend
 the GBM to a jump-diffusion by incorporating a jump process, i.e. a process of the form:
\begin{equation*}
\text{d}S_t = \mu S_t \text{d}t + \sigma S_t\text{d}B_t + S_t \text{d}J_t, 
\quad J_t = \sum_{i=i}^{N_t} Y_i
\end{equation*}
where $ J_t $ is a time-homogeneous compound Poisson process, whose jump sizes 
$ (Y_i)_{i\in\mathbb{N}} $ are a family of independently and identically distributed (iid)
random variables.\footnote{Cf. \cite{Gatheral}}

\section{A Jump-Diffusion Setting}
Lead by the observation of a volatility smile and the intuition that far-off
limit orders are underweighted, in the sense that in the geometric Brownian motion (GBM) 
framework high-level orders
are too difficult to hit, we extend our asset price model to
include jumps. In this way we will be able to account for the observed empirical phenomena and serve
intuition. Due to the increased flexibility of the models 
 we are also certain to obtain a better fit of a period-wise constant impatience 
 rate to the market data. 

 Since the success of the Black-Scholes-Merton formula for equity option valuation, which
 underlies a geometric Brownian motion, a number of jump-diffusion models have been proposed
 as an extension to the original model. We are going to use the 
\textit{Double Exponential Jump Diffusion (DEJD)} model, as proposed by Kou in \cite{Kou_OPT}
 and extensively studied by Kou and Wang in \cite{Kou_Wang}. Our choice is lead by our specific interest of 
 the Laplace transform of the first passage time of the asset price. As is 
 demonstrated in \cite{Kou_Wang}, the model offers a closed-form solution 
 (up to finding the zeros of a rational function) for the Laplace transform.

 The DEJD model of the asset mid-price $ (S_t)_{t\ge 0} $ is specified by the stochastic differential
 equation
 \begin{equation}\label{eq:dejd}
\text{d}S_t = S_t \mu \text{d}t + S_t \sigma \text{d}B_t + S_t \text{d}\left(\sum_{i=1}^{N_t}(V_i-1)\right)
\end{equation}
where $ N_t \sim \textsf{Poi(}\lambda t \textsf{)} $ is a Poisson process with intensity $ \lambda $, and
\begin{equation}
Y:=\log(V_1) \sim \textsf{dexp}(p, q, \eta_1, \eta_2\textsf{)} 
\label{eq:dexp_distro}
\end{equation}
iid random variables, distributed according to the asymmetric double-exponential law with the density
\begin{gather}
f_{Y}(t) = p\eta_1 \exp(-\eta_1 t) \textbf{1}_{\lbrace t \ge 0 \rbrace} 
+ q\eta_2 \exp(-\eta_2 t) \textbf{1}_{\lbrace t < 0 \rbrace} \label{eq:dejd_density} \\
\eta_1 > 2, \; \eta_2 > 2, \nonumber \\
p, \; q > 0, \quad p+q=1 \nonumber
\end{gather}
The model parameters can be understood as follows: at any time $ p $ is the probability of an upward jump,
and $ q= 1-p $ is the probability of a downward jump. The mean jump-sizes are $ 1/ \eta_1 $ and 
$ 1/ \eta_2 $ for an up-jump and a down-jump respectively. At each point in time only one jump
 can occur, and the occurrence of jumps is modelled by a homogeneous Poisson process with a constant  
 intensity rate of $ \lambda $, meaning that the mean number of jumps up to time $ t $ is 
 $  \lambda t $. All driving process of the model price, $ N(t) $, $ B_t $ and $ (V_i)_{i\in \mathbb{N}} $
 are assumed to be independent, although in \cite{Kou_OPT} it is suggested that this assumption can be relaxed.
 For the purposes of this paper this possibility will not be further investigated.
\begin{my_th}\label{theorem2}
The solution of the SDE \eqref{eq:dejd} is given by the process
\begin{equation}\label{eq:dejd_process}
S_t = S \exp\left(\left(\mu - \dfrac{\sigma^2}{2}\right)t + \sigma B_t \right) \prod_{i=1}^{N(t)}V_i
\end{equation}
\end{my_th}
\begin{proof} See the appendix. \end{proof}

For convenience we adopt the following representation of $ Y $:
\begin{equation}\label{eq:jump_part}
Y \stackrel{\mathcal{D}}{=} U \xi^{+} - (1-U)\xi^{-}
\end{equation}
where $ U \sim \textsf{Ber(}p\textsf{)} $ is a Bernoulli random variable, indicating that an up-jump
occurs with probability $ p $, $ \xi^{+} $ and $ \xi^{-} $ are both exponential random variables
with means $ 1/ \eta_1 $ and $ 1/ \eta_2 $ respectively, corresponding to the mean 
up-jump and mean down-jump sizes. All three random variables are assumed to be 
independent.

\begin{my_th}\label{theorem3}
 The asset mid-price model specified by \eqref{eq:dejd_process} 
 and \eqref{eq:jump_part} where $ Y=\log(V_1) $ has the following properties
\begin{itemize}
\item[a)] $ \mathbb{E}Y = \frac{p}{\eta_1} - \frac{q}{\eta_2} $ and 
 $ \mathbb{V}Y = pq\left(\frac{1}{\eta_1}+ \frac{1}{\eta_2}\right)^2
 + \left(\frac{p}{\eta_1^2} + \frac{q}{\eta_2^2} \right) $;
\item[b)] 
$
 \mathbb{E}[V] = q \frac{\eta_2}{\eta_2+1} + p \frac{\eta_1}{\eta_1-1}, 
 $
and \\
$ \mathbb{V}[V]  = p\dfrac{\eta_1}{\eta_1-2} + q \dfrac{\eta_2}{\eta_2+2} - 
\left( p\dfrac{\eta_1}{\eta_1-1} + q \dfrac{\eta_2}{\eta_2+1} \right)^2 
$,  $ \eta_1, \eta_2 > 2 $;
\item[c)] The first two moments of the ``Poisson product" $ P_t :=\prod_{i=1}^{N_t}V_i $
are:
\begin{itemize}
\item[$ \bullet $] $ \mathbb{E}[P_t] = \exp(t\lambda (\mathbb{E}[V]-1)) $ and
\item[$ \bullet $] $ \mathbb{E}[P_t^2] = \exp(t\lambda (\mathbb{E}[V^2]-1)) $;
\end{itemize}
\item[d)] 
$\mathbb{E}S_t = S \exp(\mu t)\mathbb{E}[P_t] $
and 
$\mathbb{V}S_t=S^2 e^{2\mu t}\left(e^{\sigma^2 t}\mathbb{E}[P_t^2]-\mathbb{E}[P_t]^2\right)$
\item[e)] For $ l(\Delta t) $, the log-returns over a period $ \Delta t $, we have:
\begin{itemize}
\item[$\bullet$] $ \mathbb{E}[l(\Delta t)] = \left(\mu - \frac{\sigma^2}{2}\right) \Delta t
+ \lambda\left(\frac{p}{\eta_1}- \frac{q}{\eta_2} \right) \Delta t \quad $ and
\item[$\bullet$] $ \mathbb{V}[l(\Delta t)] = \sigma^2 \Delta t + \lambda\Delta t \left(\frac{2p}{\eta_1^2} 
+ \frac{2q}{\eta_2^2} \right) $.
\end{itemize}
\end{itemize}
\end{my_th}

\begin{proof} See the appendix. \end{proof}

Observing that $ \mathbb{V}[V]=\infty $ for $ \eta_1\le 2 $,
as shown in the proof of b), we are lead to impose this constraint when modelling
 the asset price. Since there is no intuitive reason to prefer a priori any price direction
 we make the same assumption for $ \eta_2 $ in \eqref{eq:dejd_density}. 
 It essentially means, that mean jump sizes cannot exceed 50\%.

\subsection{First Passage Time Results}
Here we present the results of Kou/Wang for the first passages times of a double exponential 
jump diffusion process, as given in \cite{Kou_Wang}. Consider the DEJD proces
\begin{equation}\label{eq:kou_dejd}
X_t = \sigma B_t + \mu t + \sum_{i=1}^{N_t} Y_i; \quad X_0:=0
\end{equation}
where $ (B_t)_{t\ge 0} $ is a standard Brownian motion, $ (N_t)_{t\ge 0} $ is a Poisson process with 
intensity rate $ \lambda $, $ \mu $ and $ \sigma > 0 $ are respectively the constant drift and the volatility
of the diffusion part of the process. The family of the jump-sizes $ (Y_i)_{i\in\mathbb{N}} $ is independent
and identically distributed according to a a double-exponential distribution with density $ f_Y $ as given
in \eqref{eq:dejd_density}.
We are interested in the Laplace transform of $ \tau_b $, the random variable specified by the 
first passage time of a boundary $ b $:
\begin{equation}
\tau_b := \inf_{t\ge 0} \lbrace X_t \ge b \rbrace, \quad b > 0
\end{equation}
The infinitesimal generator of the jump diffusion process \eqref{eq:kou_dejd} is given by
\begin{equation}\label{eq:generator}
\mathcal{L}u(x)= \dfrac{1}{2}\sigma^2 u''(x) + \mu u'(x)+ \lambda 
\int_{-\infty}^{\infty}\big(u(x+y)-u(x)\big)f_Y(y) \text{d}y
\end{equation}
for all twice continuously differentiable functions $ u(x) $. Further, suppose $ \theta \in ]-\eta_2,
\eta_1[ $. The moment generating function of the jump size $ Y $ is given by:
\begin{equation}
\mathbb{E}[ \exp(\theta Y )] = \dfrac{p\eta_1}{\eta_1 - \theta} + \dfrac{q\eta_2}{\eta_2+\theta},
\end{equation}
from which the moment generating function of $ X_t $ can be obtained as 
\begin{equation*}
\varphi(\theta, t) :=\mathbb{E}[\exp(\theta X_t )]=\exp(G(\theta)t),
\end{equation*}
where the function $ G(\cdot) $ is defined as
\begin{equation}
G(x) := x\mu + \dfrac{1}{2}x^2\sigma^2 + \lambda
\left( \dfrac{p\eta_1}{\eta_1 - x} + \dfrac{q\eta_2}{\eta_2+x} - 1 \right)
\end{equation}

\begin{my_lem}\label{lemma_G}
The equation \begin{equation}
G(x) = \alpha \; \text{ for all } \; \alpha >0
\label{eq:func_G}
\end{equation} 
has exactly four roots: $ \beta_{1,\alpha} $, $ \beta_{2,\alpha} $, $ -\beta_{3,\alpha} $, 
$-\beta_{4,\alpha} $ where
\begin{equation*}
0 < \beta_{1,\alpha} < \eta_1 < \beta_{2,\alpha} < \infty, \quad 
0 < \beta_{3,\alpha} < \eta_2 < \beta_{4,\alpha} < \infty
\end{equation*}
\end{my_lem}

\begin{proof} Cf. \cite{Kou_Wang} (page 507, Lemma 2.1). \end{proof}

\begin{my_th}\label{theorem4}
For any $ \alpha \in ]0, +\infty[ $, leta $ \beta_{1,\alpha} $ and $ \beta_{2,\alpha} $
be the only positive roots of the equation
\begin{equation*}
\alpha = G(x),
\end{equation*} 
where $ 0 < \beta_{1,\alpha} < \eta_1 < \beta_{2,\alpha} < + \infty $. Then the Laplace transform
of $ \tau_b $ is given by:
\begin{equation}\label{eq:eq_theorem4}
\mathbb{E}\left[e^{-\alpha \tau_b}\right] = \dfrac{\eta_1 -\beta_{1,\alpha}}{\eta_1}
\dfrac{\beta_{2,\alpha}}{\beta_{2,\alpha}-\beta_{1,\alpha}}e^{-b\beta_{1,\alpha}}
+ \dfrac{\beta_{2,\alpha}-\eta_1}{\eta_1} 
\dfrac{\beta_{1,\alpha}}{\beta_{2,\alpha}-\beta_{1,\alpha}}e^{-b\beta_{2,\alpha}}
\end{equation}
\end{my_th}

\begin{proof} Cf. \cite{Kou_Wang} (page 509, Theorem 3.1). \end{proof}

Again, a simple transform is needed in order to apply the result from the previous
theorem to our process given by \eqref{eq:dejd_process}. For $ D>0 $ consider:

\begin{align}
\tau_{S+D}^{\text{exp}} & =\inf_{t\ge 0}\lbrace S_t \ge S + D \rbrace \nonumber \\
& =\inf_{t\ge 0}\left\lbrace S \exp\left(\left(\mu - \dfrac{\sigma^2}{2}\right)t
+ \sigma B_t + \sum_{i=1}^{N_t}Y_i \right) \ge S + D \right\rbrace \nonumber \\
& = \inf_{t\ge 0} \Big\lbrace 
\underbrace{ \left(\mu-\dfrac{\sigma^2}{2}\right)}_{\hat{\mu}} t 
 +\sigma B_t + \sum_{i=1}^{N_t} Y_i \ge \underbrace{\log\left(\dfrac{S+D}{S}\right)}_{z} \Big\rbrace \nonumber\\
& = \hat{\tau}_z
\label{eq:dejd_ask_passage}
\end{align}

So the expected utility for the ask side of the LOB (i.e. for the sellers) is given by
substituting $ z $ for $ b $ in \eqref{eq:eq_theorem4}, and  calculating 
$ \beta_{1,r} $ and $ \beta_{2,r} $ by solving \eqref{eq:func_G} with $ \hat{\mu} $ and $ z $:
\begin{gather*}
G(\beta_i) = r, \quad i\in\lbrace 1,2 \rbrace \\
\beta_i \hat{\mu} + \dfrac{1}{2}\beta_i^2\sigma^2 + \lambda
\left( \dfrac{p\eta_1}{\eta_1 - \beta_i} + \dfrac{q\eta_2}{\eta_2+\beta_i} - 1 \right) = r
\intertext{where the two roots $ \beta_1 $ and $ \beta_2 $ are in the following intervals:}
0 < \beta_{1,\alpha} < \eta_1 < \beta_{2,\alpha} < \infty
\end{gather*}

Observe, that while in the GBM setting, we had direct access to a formula for the 
first hitting time regardless of whether the boundary was above, or below the starting
point of the process, Theorem \ref{theorem4} only provides a formula for higher boundaries.
So we need to make a distinct transform for the bid (i.e. buy) side of the LOB. Now, consider
for $ D>0 $ the following:
\begin{align}
\tau_{S-D} & = \inf_{t\ge 0}\lbrace S_t \le S - D \rbrace \nonumber\\
& = \inf_{t\ge 0} \Bigg\lbrace \underbrace{-\left(\mu-\dfrac{\sigma^2}{2}\right)}_{\hat{\mu}}t 
 -\sigma B_t - \sum_{i=1}^{N_t} Y_i \ge 
 \underbrace{-\log\left(\dfrac{S-D}{S}\right)}_{z} \Bigg\rbrace \nonumber\\
 & \stackrel{\mathcal{D}}{=} \inf_{t\ge 0} \left\lbrace 
\hat{\mu}  t 
+ \sigma B_t - \sum_{i=1}^{N_t} Y_i \ge  z \right\rbrace \nonumber\\
 &  \stackrel{\mathcal{D}}{=} \inf_{t\ge 0} \left\lbrace 
\hat{\mu}t 
 + \sigma B_t + \sum_{i=1}^{N_t} \hat{Y_i} \ge z \right\rbrace \nonumber\\
 & = \hat{\tau}_{z}
\label{eq:dejd_bid_passage}
\end{align}
where the first transform in probability is due to the reflection 
principle\footnote{Cf. \cite{Klenke}.}, and the second due to 
\begin{gather}
Y=U\xi^{+} - (1-U)\xi^{-} \Rightarrow -Y = -(1-U)\xi^{-}-U\xi^{+}
\intertext{where $ Y\sim\textsf{dexp(}p,q,\eta_1, \eta_2 $), so that for $ \hat{Y}\sim\textsf{dexp(} q,p,\eta_2,\eta_1$)}
\hat{Y}\stackrel{\mathcal{D}}{=}-Y.
\end{gather}

Obviously in the DEJD model there are significantly more parameters to be 
either estimated, or fitted. While in the GBM setting we could extract the 
impatience rate and expected utility by estimating the drift and volatility of
log-returns, and imposing constraints such as keeping $ r $ and $ U_D(r) $ constant, 
a similar strategy would have only limited success in the DEJD framework. A fundamental trade-off between 
keeping the model as flexible as possible and imposing new constraints is at hand, 
because the former carries a significant risk of overfitting, and the latter
might not converge for a large class of pre-set parameters. Next we show how to 
separate the diffusive volatility from the jump-part contribution to total 
process volatility, which reduces the number of parameters needed to fit.

\subsection{Parameter Calibration to Market Data}
The main difference in this process, compared to a diffusion, is that a 
point estimate of $ \mathbb{V}[l(\Delta)t] $ from observed
data is an estimate for the sum of the diffusive and the jump-part variances,
as is shown in Theorem \ref{theorem2}. We will therefore make use of the 
bipower variation introduced by Brandorff-Nielsen and Shephard in \cite{Brandorff}. 
Consider the realized variance over a period $ N \Delta t $:\footnote{This presentation is 
based on the exposition in \cite{Thierry} with a number of alterations to better 
suite the purpose of this paper. }
\begin{equation*}
RV_{N \Delta t}\left(\frac{1}{N}\right):=\sum_{j=1}^{N}R_{j,\Delta t}^2(N)
\end{equation*}
where $ N $ is the sampling frequency and $ R_{j,\Delta t}^2(N) $ is the log return in 
the time span from $ (j-1)\Delta t $ to $ j\Delta t $. Notice that 
$ RV_{N \Delta t} $ is an estimate of the total process variance over the whole sampling period
$ N\Delta $. It can be shown\footnote{Cf. \cite{Brandorff}}, that 
\begin{equation*}
\text{for }\;  m\to \infty \; RV_{m \cdot\frac{\Delta t N }{m}}\left(\frac{1}{m}\right) \to 
\mathbb{V}[l(N\Delta t)],
\end{equation*}
i.e. by increasing the sampling frequency over the same interval $ N \Delta t$, the realized
variance converges to the the process' total variance over the period $ N\Delta t $. 
Consider further the bipower variation:
\begin{equation*}
BV_{N \Delta t}\left(\frac{1}{N}\right)=\dfrac{\pi}{(1-2/N)2}\sum_{j=3}^{N}
\left\vert R_{j, \Delta t}^2(N)\right\vert \left\vert R_{j-2, \Delta t}^2(N)\right\vert,
\end{equation*}
which is shown to converge for an increasing sampling frequency to the diffusive part of the 
total variance of the process over the whole sampling period $ N \Delta t $, i.e.
\begin{equation*}
\text{for }\;  m\to \infty \; BV_{m \cdot\frac{\Delta t N }{m}}\left(\frac{1}{m}\right) \to 
\sigma^2 N \Delta t
\end{equation*}
So from the log-returns $ l(\Delta t) $ we can estimate the diffusive 
$ \sigma^2 $ by:
\begin{equation}
\sigma^2 \Delta t = \dfrac{BV_{N \Delta t}\left(\frac{1}{N}\right)}{N}
\label{eq:sigma_BV_diff}
\end{equation}
and for the jump part we can use:
\begin{gather}
\mathbb{V}[l(\Delta t)]-\sigma^2\Delta t
= \dfrac{RV_{N \Delta t}\left(\frac{1}{N}\right)-BV_{N \Delta t}\left(\frac{1}{N}\right)}{N}, \nonumber
\intertext{which in combination with the estimate for $ \sigma^2 $ and with Theorem
\ref{theorem2} e) leads to}
\lambda \Delta t \left(\frac{2p}{\eta_1^2}+\dfrac{2q}{\eta_2^2}\right) = 
\dfrac{RV_{N \Delta t}\left(\frac{1}{N}\right)-BV_{N \Delta t}\left(\frac{1}{N}\right)}{N},
\label{eq:sigma_BV_jump}
\end{gather}
quantifying the jump-part contribution to the process' total variance. Due to 
sampling errors however, this expression may take a negative value, so in order to avoid
this in our empirical work we cap the bipower variation by the realized variance:
\begin{equation}
\widehat{BV}_{N \Delta t\left(\frac{1}{N}\right)}
:=\min\left\lbrace BV_{N \Delta t}\left(\frac{1}{N}\right), 
RV_{N \Delta t}\left(\frac{1}{N}\right)\right\rbrace
\label{eq:sigma_BV_constraint}
\end{equation}
Again, as in the GBM setting, we choose $ N=30 $ and base our rolling estimates 
at $ t $ on observed log-returns from $ t-N-1 $ to $ t-1 $.
Still, including the diffusive drift $ \mu $, we have a total of five 
parameters (as $ q=1-p $) to estimate 
from a single constraint. For this reason we assume, that the process has no drift, 
arguing that price level changes come about jump-wise, and that
price movement between jumps is driven only by Brownian motion scaled by its diffusive
$ \sigma $. We can also add another constraint from the observed rolling mean 
of the log-returns:
\begin{equation}
\overline{m}=-\dfrac{BV_{N \Delta t}\left(\frac{1}{N}\right)}{2N}
+\lambda \Delta t \left(\frac{p}{\eta_1}-\dfrac{q}{\eta_2}\right)
\label{eq:mean_BV_constraint}
\end{equation}
Now there are four free parameters and two constraints. Additional assumptions and 
constraints may include either, or all of the following (superscripts indicate
the time of the data point):
\begin{itemize}
\item [$ \bullet $] $ \eta_1^{j}=\eta_2^{j} $, or $ \left(\eta_1^{j}-\eta_2^{j}\right)^2 
\le \epsilon $, which amounts to assuming that mean jump-sizes are essentially the 
same for up-jumps and down-jumps. The only source of asymmetry in the model derives 
from the probability $ p $ that an up-jump occurs, given that the price process jumps at all.
\item [$ \bullet $] $ \left(\lambda^{j}-\lambda^{j-1}\right)^2\le \epsilon $;
\item [$ \bullet $] $ \left(p^{j}-p^{j-1}\right)^2\le \epsilon $ 
\item [$ \bullet $] $ \left(\eta_i^{j}-\eta_i^{j-1}\right)^2\le \epsilon $ for 
$ i=1,2 $.
\end{itemize}
All of these constraints make sense, and especially the first one would be the most stringent
and effective, as the estimation problem would then be to derive three parameters
from two constraints. However, readily available optimization procedures would not converge
when this constraint is introduced for reasons explained in the next paragraph. 

Another feature of the fitting is that we now need to fit the impatience rate and expected
utility for the bid and for the ask side simultaneously. While in the GBM setting
the process parameters for the bid and the ask side were the same and did not need fitting, 
which allowed us to solve for the impatience rate and the expected utility on each side of the book
consecutively, in the DEJD setting we must assume that sellers and buyers have the same view 
of the underlying process, so we need to fit two impatience parameters
$ r^{\text{ask}} $ and $ r^{\text{bid}} $ in parallel, while still under the assumption that 
either of the expected utilities $ U^{\text{bid}} $ and $ U^{\text{ask}} $ 
is step-wise approximately constant and evolves smoothly. 
This of course makes the fitting procedure much more difficult 
and potentially unstable with off-the-shelf optimization techniques. In fact, this is the 
reason why with the introduction of more stringent constraints from the above list, the 
optimization fails to converge - we have found that for a target function similar 
to the one from GBM framework minimizing the error for one side of the book only, 
an optimization algorithm with all of the above constraints introduced would converge.

On one hand, fitting both sides simultaneously restrains available optimization techniques
from introducing constraints. On the other hand fitting the stochastic process' driving parameters
on two separate and independent sets of data in parallel is of and in itself an additional 
constraint, which should discipline the optimization algorithm and provide insurance against
parameter redundancy. Since we can optimize one side of the book imposing all of the constraints, 
we suppose a more careful analysis of the parameter domain
for the optimization problem should lead to an algorithm which converges. This, however,
is beyond the scope of this paper, as it is a numerically challenging problem, which 
departs from the research of price dynamics implications of the LOB. 
Furthermore, with only the last constraint from the list above, we achieve reasonable results.

Following is a pseudo-code representation of Algorithm \ref{DEJD_algo}
for the estimation of the impatience rate in a DEJD setting.
The inputs are starting utilities and impatience rates for both sides of the book,
and a vector of process starting parameters $ P $, which cannot be directly estimated and need fitting, that is 
$ \lambda $, $ \eta_1 $, $ \eta_2 $ and $ p $. The outputs are two vectors of utilities,
two vectors of impatience rates and a set of price process parameters derived by our fitting
procedure.
\begin{algorithm}
\caption{Estimation of the impatience rate in a DEJD setting}
\label{DEJD_algo}
\textbf{Inputs:} N, steps, $ U_0^{\text{ask}} $, $ r_0^{\text{ask}} $, 
$ U_0^{\text{bid}} $, $ r_0^{\text{bid}} $, $ s^2 $, $ m $
\begin{algorithmic}[1]
\STATE $ r_0:=\big( r^{\text{ask}}_0, r_0^{\text{bid}} \big) \in \mathbb{R}^2_{+} $
\STATE $ U_0:=\big( U^{\text{ask}}_0, U_0^{\text{bid}} \big) \in \mathbb{R}^2_{+} $
\FOR {i=1:N } 
	\WHILE{$ \text{error(r)}>\epsilon \; \& \; \text{Tolerance} > \delta $}
		\STATE $ r_i := \min_{r>0.01}\lbrace \text{error(r)} \rbrace $ 
		\STATE s.t. constraints \eqref{eq:sigma_BV_diff},\eqref{eq:sigma_BV_jump},\eqref{eq:sigma_BV_constraint},
\eqref{eq:mean_BV_constraint} \&  $ \left(\eta_i^{j}-\eta_i^{j-1}\right)^2\le \epsilon $ for $ i=1,2 $
		\STATE \textbf{WHERE} error(r) is 
		\STATE	calculate $ U_j^{\text{ask}} $ according to \eqref{eq:the_utility}, \eqref{eq:dejd_ask_passage}, using $ P $, $ r^{\text{ask}} $
		\STATE	calculate $ U_j^{\text{bid}} $ according to \eqref{eq:the_utility}, \eqref{eq:dejd_bid_passage}, using $ P $, $ r^{\text{bid}} $
		\STATE error(r) = $ \left( \dfrac{(U^{\text{ask}}_j - U^{\text{ask}}_{j-1})}						 						{0.5(U^{\text{ask}}_j + U^{\text{ask}}_{j-1})} \right)^2 +  		 			\left( \dfrac{(U^{\text{bid}}_j - U^{\text{bid}}_{j-1})}						 						{0.5(U^{\text{bid}}_j + U^{\text{bid}}_{j-1})} \right)^2 $
		\STATE error(r) = $ \left( \dfrac{(r^{\text{ask}}_j - r^{\text{ask}}_{j-1})}						 						{0.5(r^{\text{ask}}_j + r^{\text{ask}}_{j-1})} \right)^2 + 				 		\left( \dfrac{(r^{\text{bid}}_j - r^{\text{bid}}_{j-1})}						 						{0.5(r^{\text{bid}}_j + r^{\text{bid}}_{j-1})} \right)^2 $					
		\STATE error(r) =error(r)/4									
	\ENDWHILE
\ENDFOR
\end{algorithmic}
\textbf{Outputs:} $ U =\big(U^{\text{ask}}, U^{\text{bid}}\big)\in \mathbb{R}^{2\times N}_{+}$,
 $ r = \big(r^{\text{ask}}, r^{\text{bid}}\big)\in \mathbb{R}^{2\times N}_{+}$,
 $ P \in \mathbb{R}^{4\times N} $
\end{algorithm}

As already noted, this problem is numerically challenging. Very sensible tuning of 
optimization parameters concerning convergence tolerance is needed. The problem is 
further complicated, by the fact that the function $ G $ (cf. Theorem \ref{theorem4}) 
is so poorly conditioned, that finding its roots in the specified intervals (where 
the function is differentiable) becomes very hard. For this reason we must use an efficient 
implementation of the very robust but time-consuming bisection\footnote{Cf. 
\cite{Burden}} algorithm. Note that in order to apply bisection the function $ G $ needs to 
be monotonic and continuous over the specified interval, and its values at the 
interval boundaries to have different signs. This is verified in Lemma \ref{lemma_G}.

\section{Results For a Jump-Diffusion}
\begin{figure}[!ht]\centering
\includegraphics[scale=0.9]{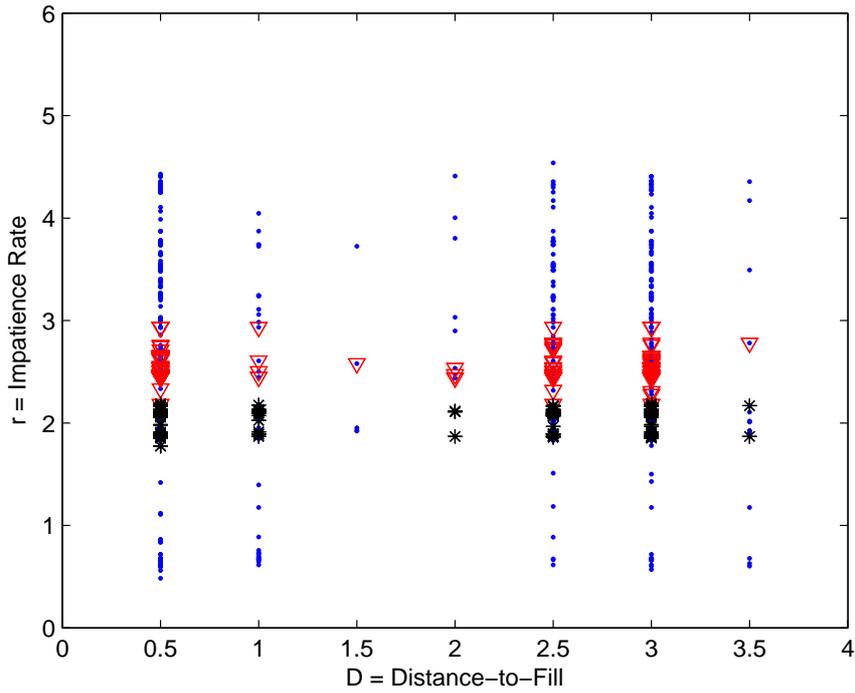}
\caption{$ r $ against $ D $. Apparently much less noisy than the GBM
framework, but at first sight still very scattered. Notice, however the black 
stars ($ r $ from $ t=$3 to $ t=$3.75) and the red triangles ($ r $ from $ t=$5.25 to $ t=$5.75),
as a function of $ D $ in two different market modes.
In the absence of market sentiment shifts $ r $ is nearly
constant. Changes in $ r $ could indicate shift in sentiment. 
It is also apparent that in contrast to the GBM setting,
the in the DEJD model impatience rate is a function of neither the distance-to-fill, nor the volatility.}
\label{fig:DEJD_D_r}
\end{figure}

The double-exponential jump-diffusion (DEJD) proves to be much better suited to model 
an equilibrium in the LOB 
under the assumption of a constant impatience rate across levels, given an established 
market regime (see Figure \ref{fig:DEJD_D_r}). The induced fat-tailed distribution of log-returns 
makes orders further out in the book more accessible, than in a GBM setting. We further establish, that
in order to achieve an equilibrium of the LOB, as reflected by the consistency of the impatience rate, the underlying process cannot be e GBM.

\begin{figure}[!ht]\centering
\includegraphics[scale=0.9]{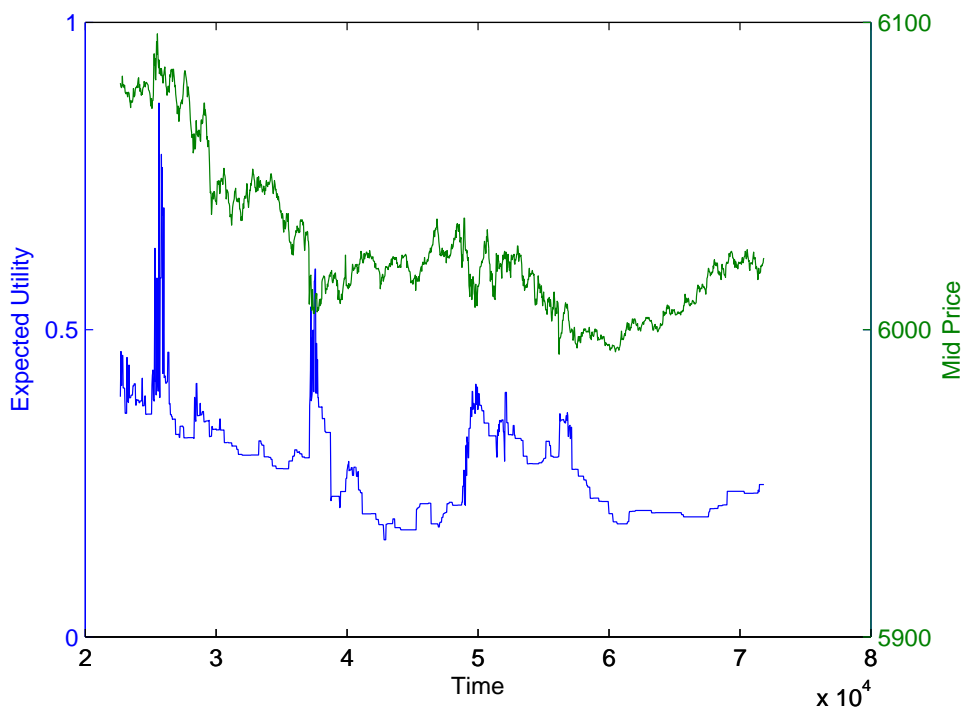}
\caption{Expected utility (left axis) and mid-price (right axis) for the sell side
of the book in the DEJD setting. The utility is not as smooth as in the GBM
setting, but still a very good fit. (FDAX, August 20, 2010)}
\label{fig:DEJD-U-Mid}
\end{figure}

\begin{figure}[!ht]\centering
\includegraphics[scale=0.9]{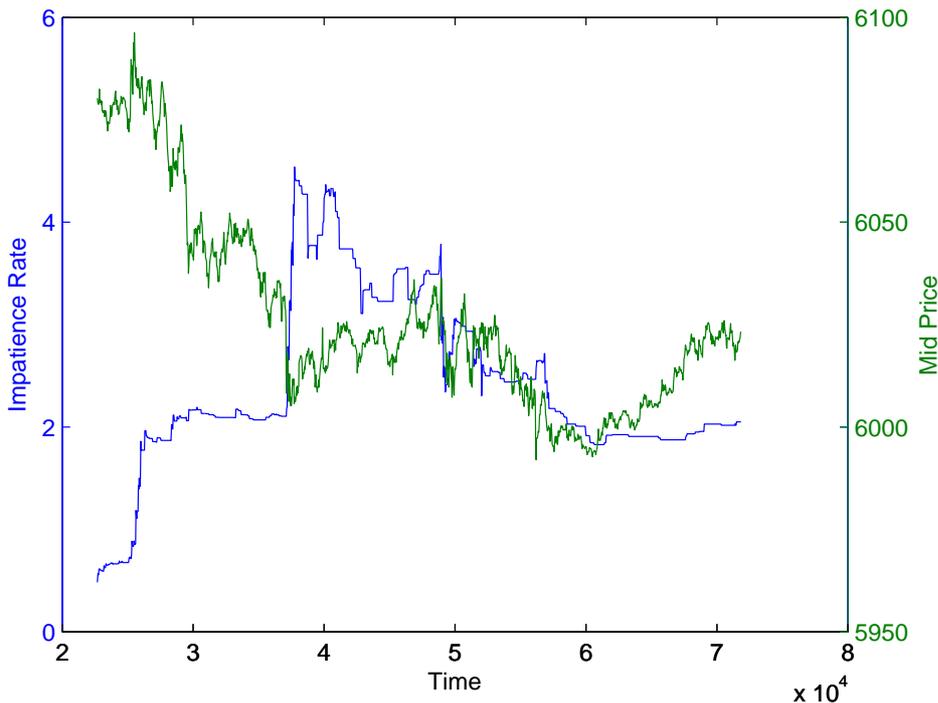}
\caption{Impatience rate (left axis) and mid-price (right axis).
The DEJD framework provides for a smooth, level-independent $ r $, where the GBM setting fails.
(FDAX, August 20, 2010)}
\label{fig:DEJD-r-Mid}
\end{figure}

We illustrate our findings by showing a representative day, again August 20, 2010, and again only
showing results for the sell-side of the LOB. Figure \ref{fig:DEJD-U-Mid}
shows the expected utility in the DEJD setting. While still essentially flat for established market regimes,
its choppiness is indicative of partial numerical instability, a problem outlined in the previous section. 
Nonetheless, the success of the DEJD over the GBM is clearly demonstrated 
in Figure \ref{fig:DEJD-r-Mid}, where the evolution
of the impatience rate is shown. It is very stable for established market modes and changes spike-wise
to indicate that a new market regime has been established. Regression analysis has however showed,
that as in the GBM case, the impatience rate is not indicative of future price movements. It 
serves very much as an anchor which allows for consistent comparison of limit orders at different levels
in the LOB, as long as market sentiment remains unchanged.

Figure \ref{fig:DEJD_D_r} clearly shows, that in a DEJD setting, an approximately constant impatience rate 
can be derived. It shows the impatience rate as a function of the distance-to-fill in the course of a
whole day. Highlighted are the values for two periods when market regime was unchanged, 
and the expected utility was constant in each period. Since the implied impatience rates form a flat line across all 
distances for a given level of utility, they are not dependent on the limit order level. This
means that $ r $ can be used as reliable and well-defined quantifier of the trade-off between
waiting longer for a better price against waiting less for a worse price.

\section{Conclusion}
We have shown that the assumption of an equilibrium in the limit order book (LOB) 
is not consistent with the price dynamics
given by a geometric Brownian motion (GBM). 
Instead, in equilibrium, the LOB implies a volatility smile that is reminiscent of, 
but unrelated to, the one 
known from option-pricing. This observation necessarily leads
to a fat-tailed distribution of log-returns.\footnote{In 
\cite{Kou_OPT} it is shown, that this is the case in the DEJD model.} The most
natural explanation of this phenomenon is the occurrence of jump processes in the price dynamics.

We further demonstrate how an impatience rate is implied from empirical observations,
so that it is consistent with the assumption of market efficiency. 

There are several directions for further research of the proposed framework.
A natural extension would be the study of different jump-diffusion models, which provide greater
parameter stability. In addition one could investigate
stochastic volatility models as an alternative to incorporating the volatility smile. 
Furthermore it would be instructive to
compare the jump parameters presented in this paper with the results from standard techniques 
that are based on a direct analysis of the time-series.

\section{Acknowledgement}
We would like to thank Gregor Svindland for constructive comments on this paper.

\appendix

\section{Additional Plots}

\begin{figure}[H]
  \centering
  \subfloat[GBM Buyer]{\includegraphics[scale=0.47]{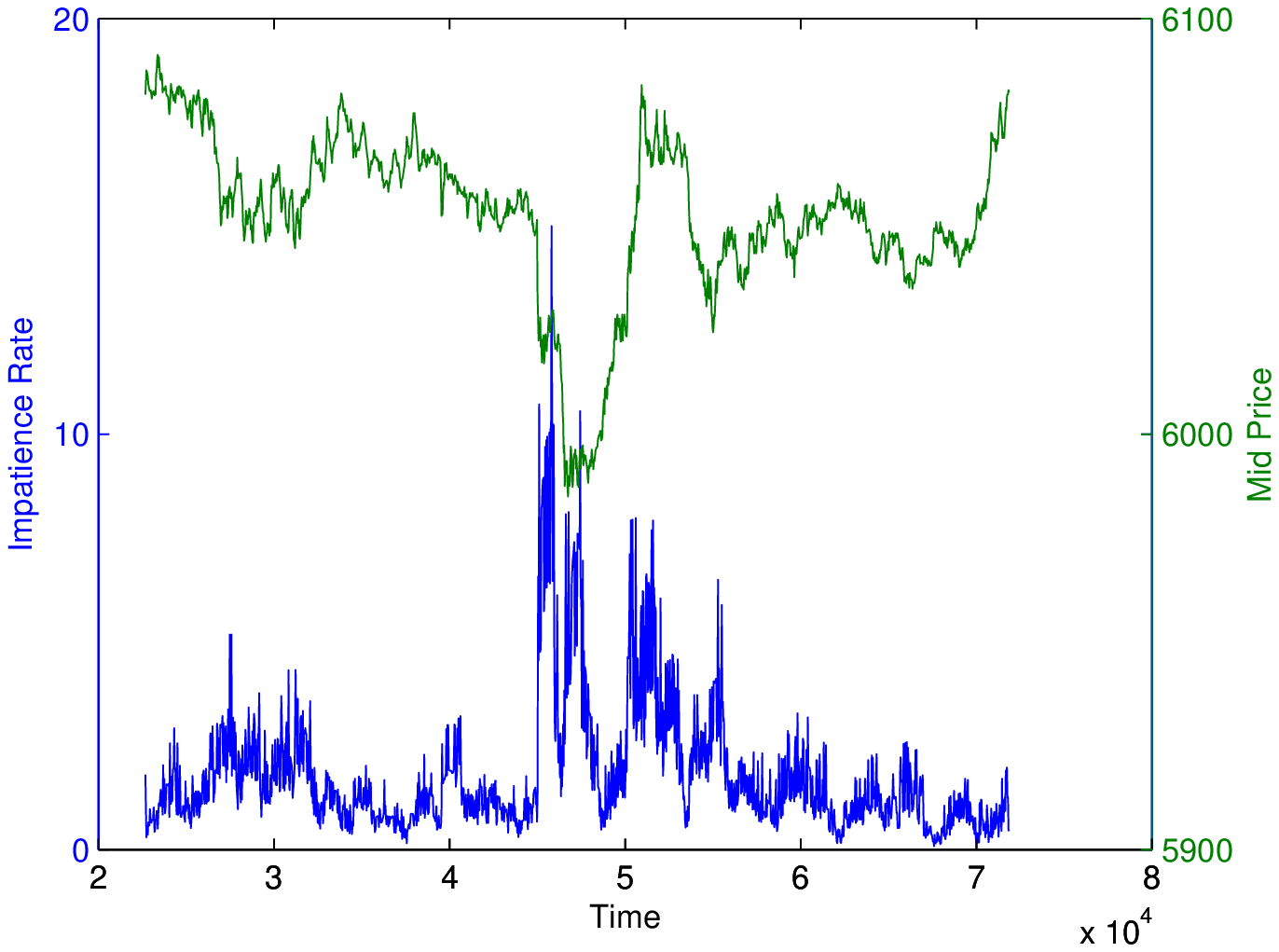}}                
  \subfloat[DEJD Buyer]{\includegraphics[scale=0.47]{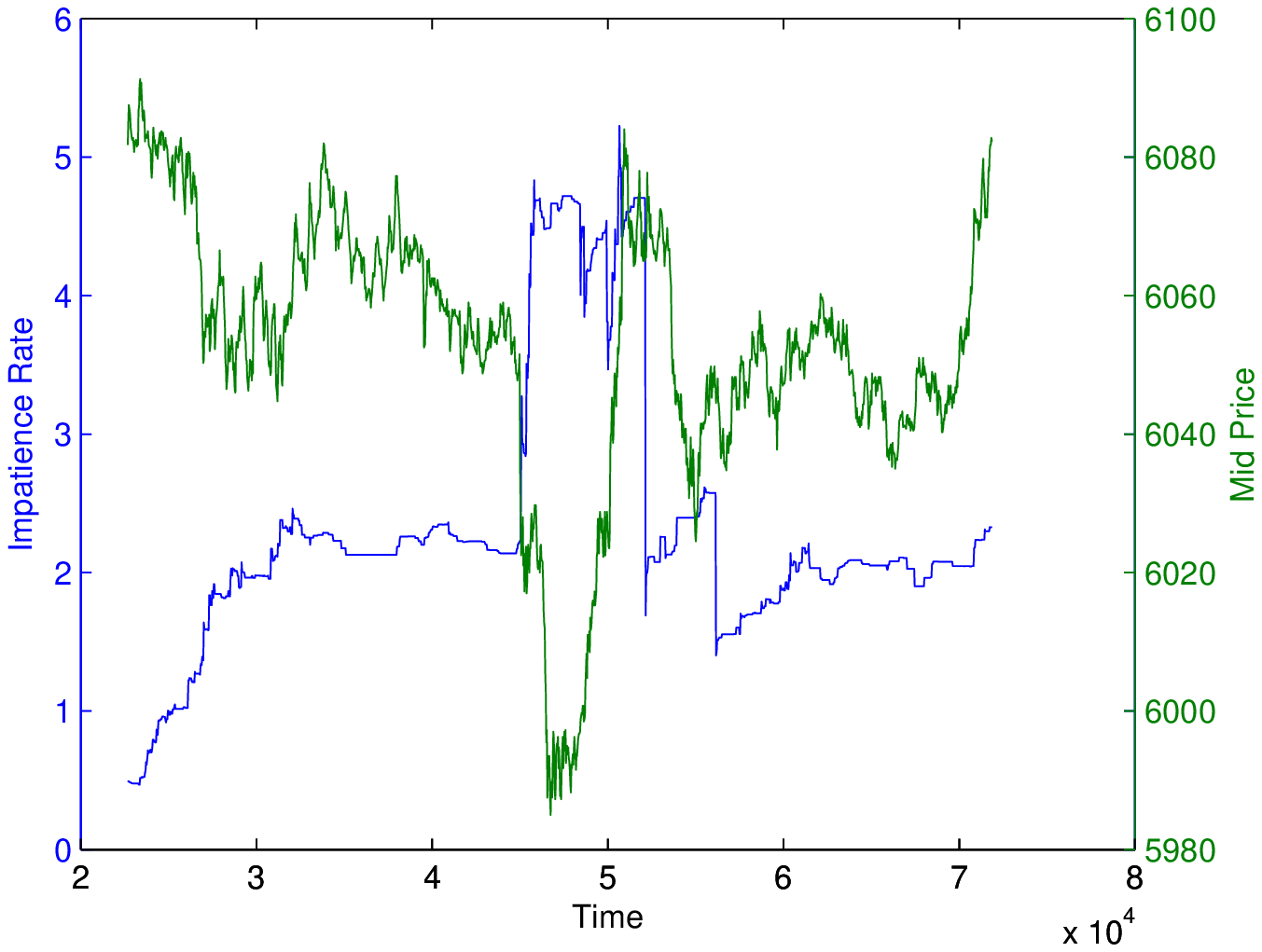}}
\end{figure}
\begin{figure}[H]
  \centering
  \subfloat[GBM Seller]{\includegraphics[scale=0.47]{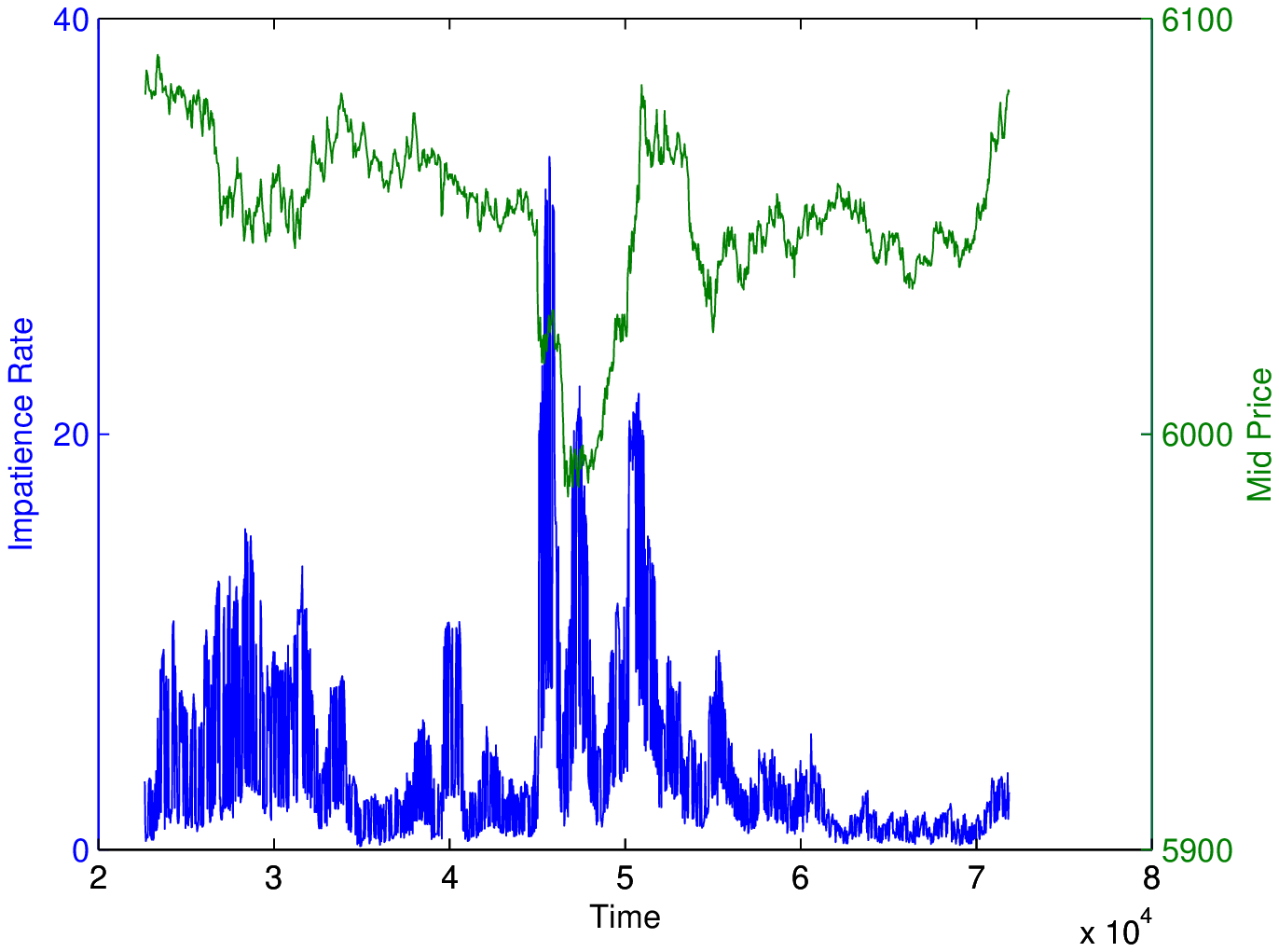}}                
  \subfloat[DEJD Seller]{\includegraphics[scale=0.47]{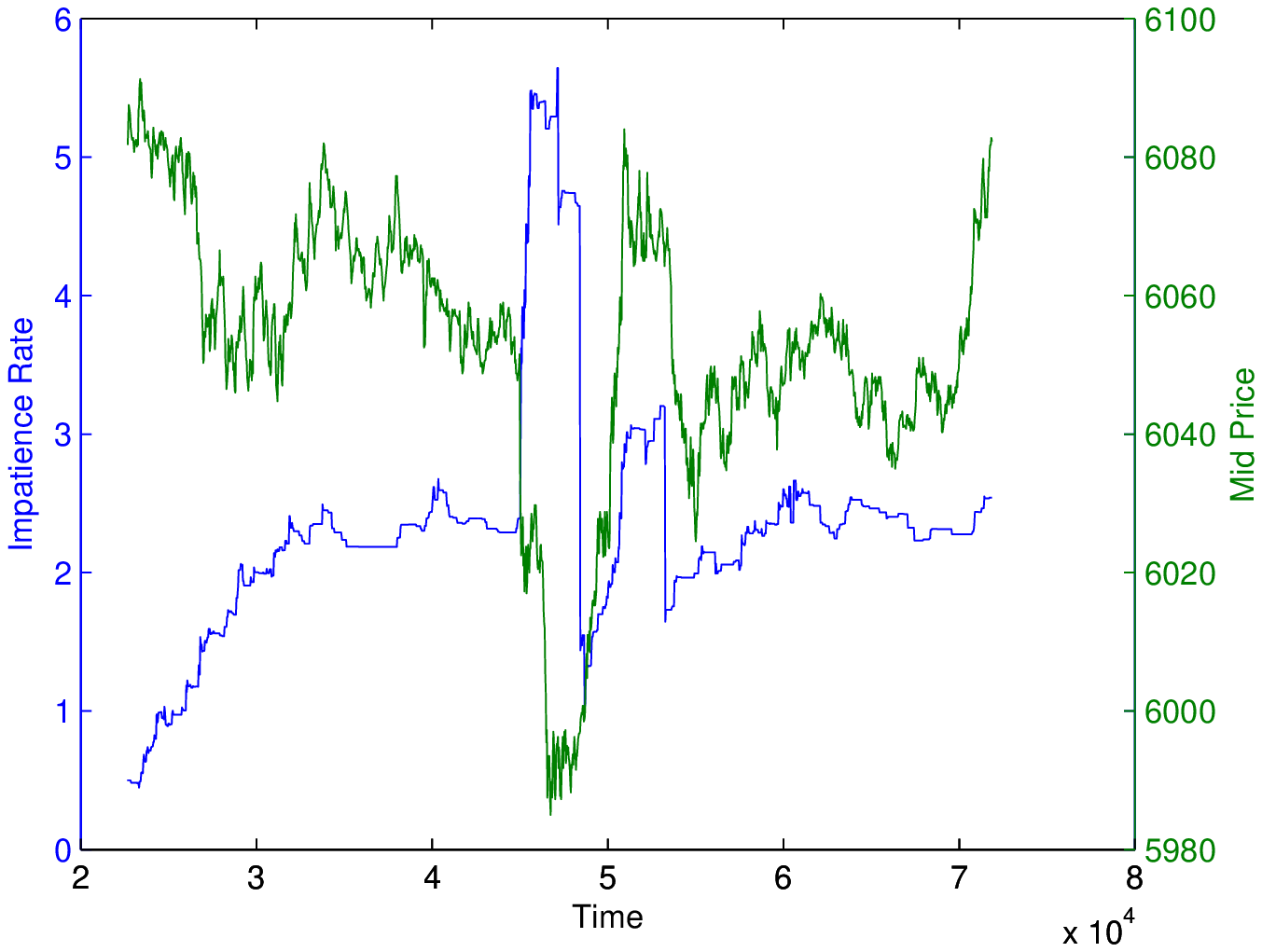}}
  \caption{Impatience rates and mid-price.}
\end{figure}

\begin{figure}[H]
  \centering
  \subfloat[GBM Buyer]{\includegraphics[scale=0.47]{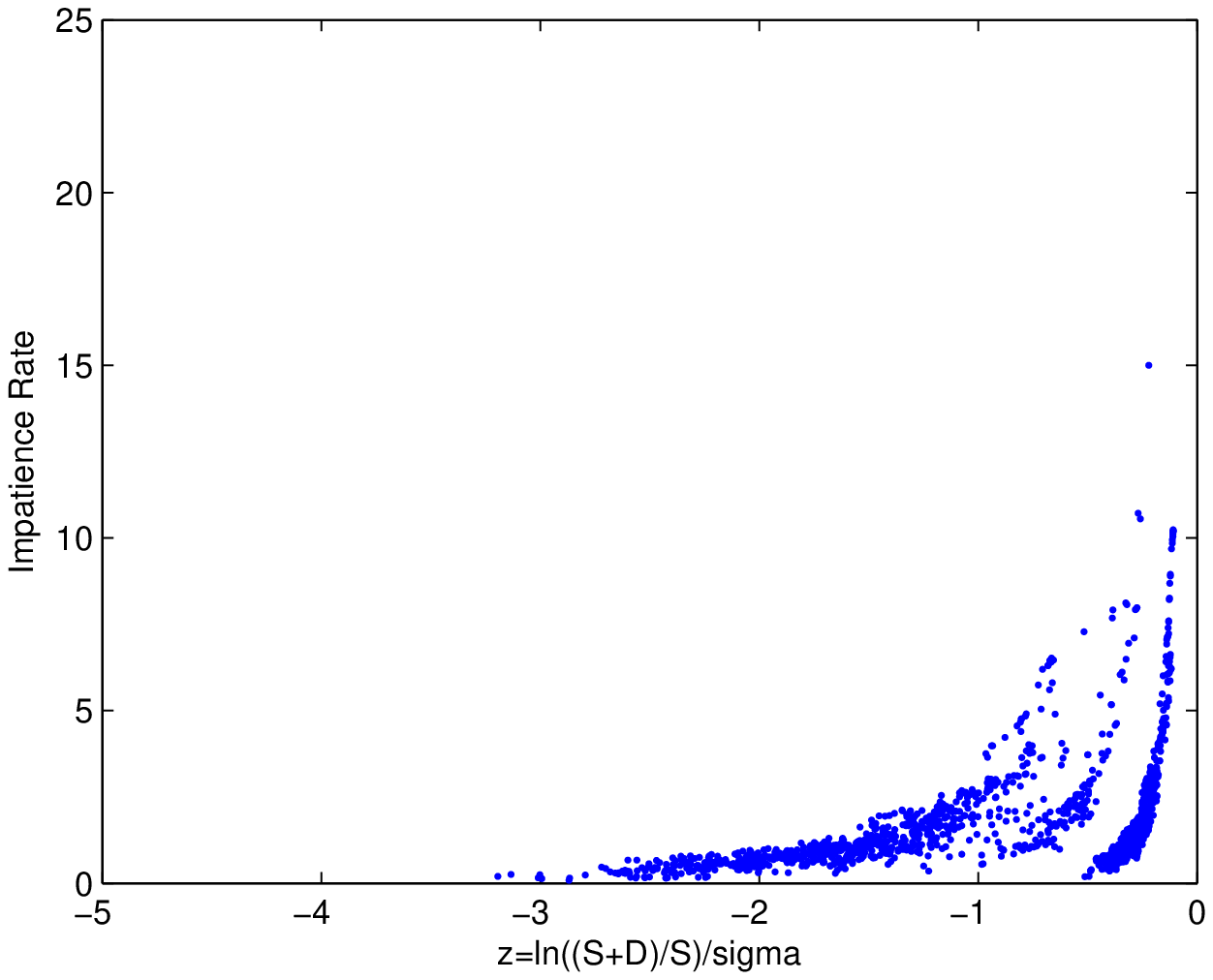}}                
  \subfloat[DEJD Buyer]{\includegraphics[scale=0.47]{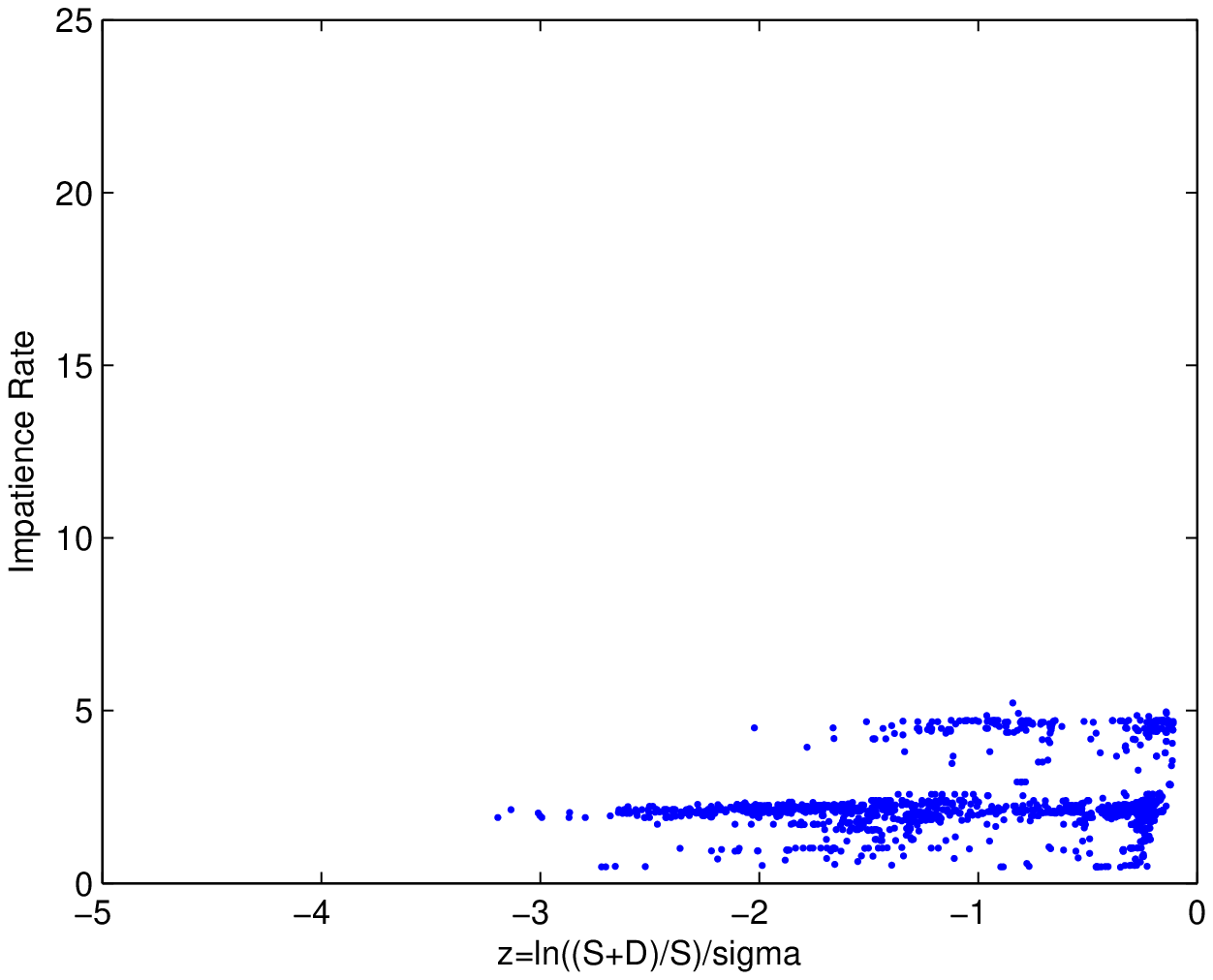}}
\end{figure}
\begin{figure}[H]
  \centering
  \subfloat[GBM Seller]{\includegraphics[scale=0.47]{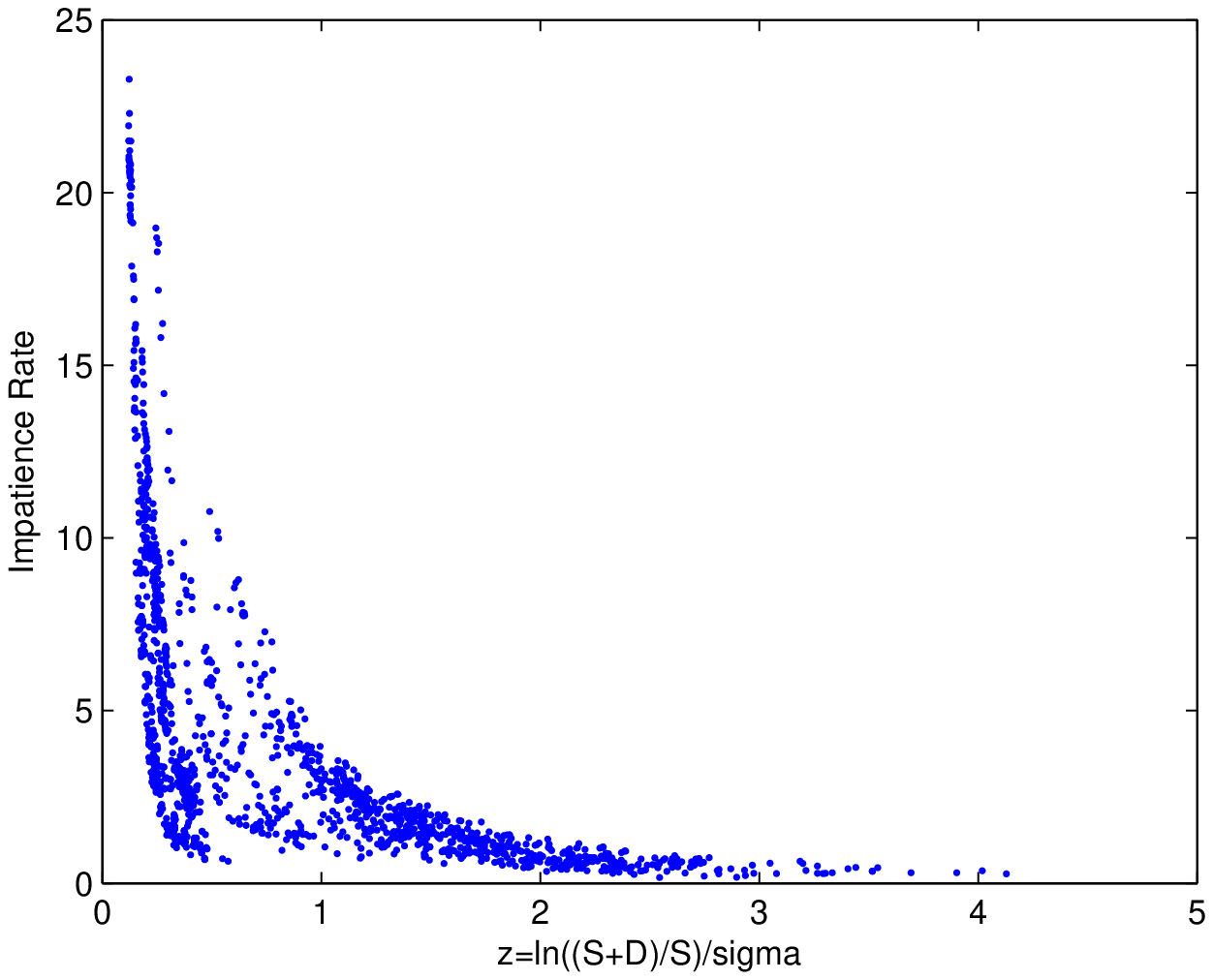}}                
  \subfloat[DEJD Seller]{\includegraphics[scale=0.47]{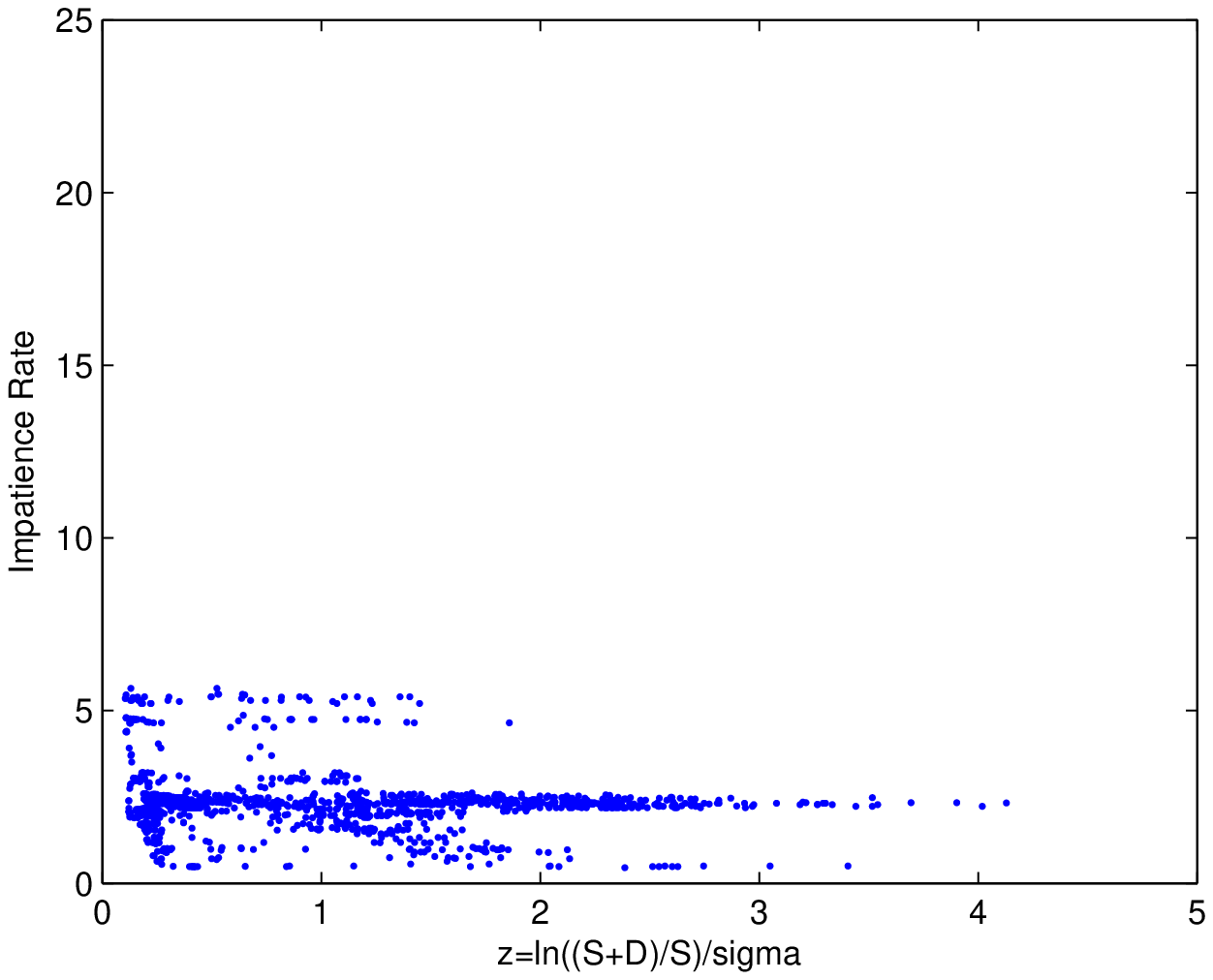}}
  \caption{Impatience rates against the risk-adjusted realtive distance-to-fill
  $ z=\frac{\log(S+D)/S}{\sigma} $.}
\end{figure}

\begin{figure}[H]
  \centering
  \subfloat[GBM Buyer]{\includegraphics[scale=0.47]{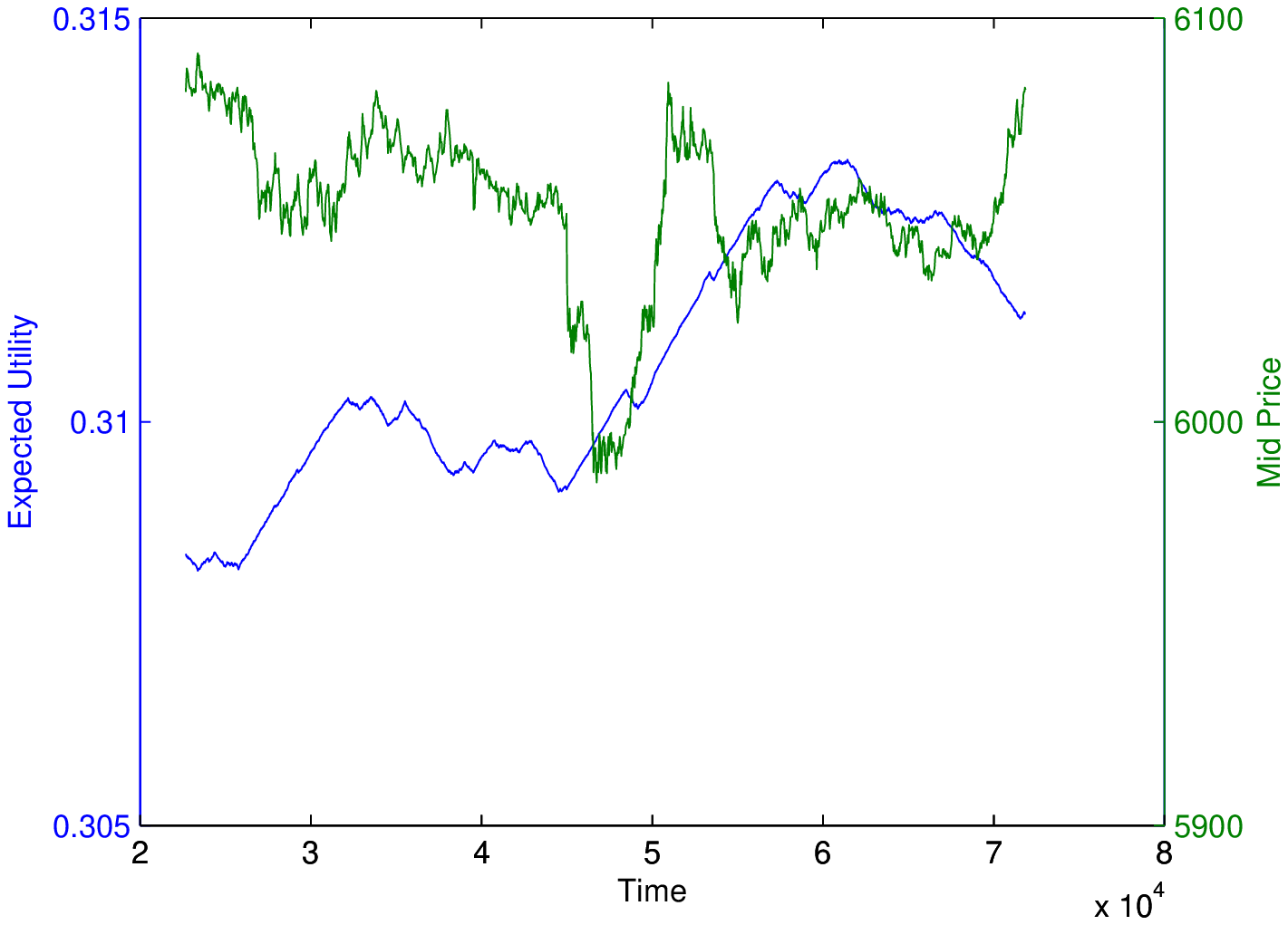}}                
  \subfloat[DEJD Buyer]{\includegraphics[scale=0.47]{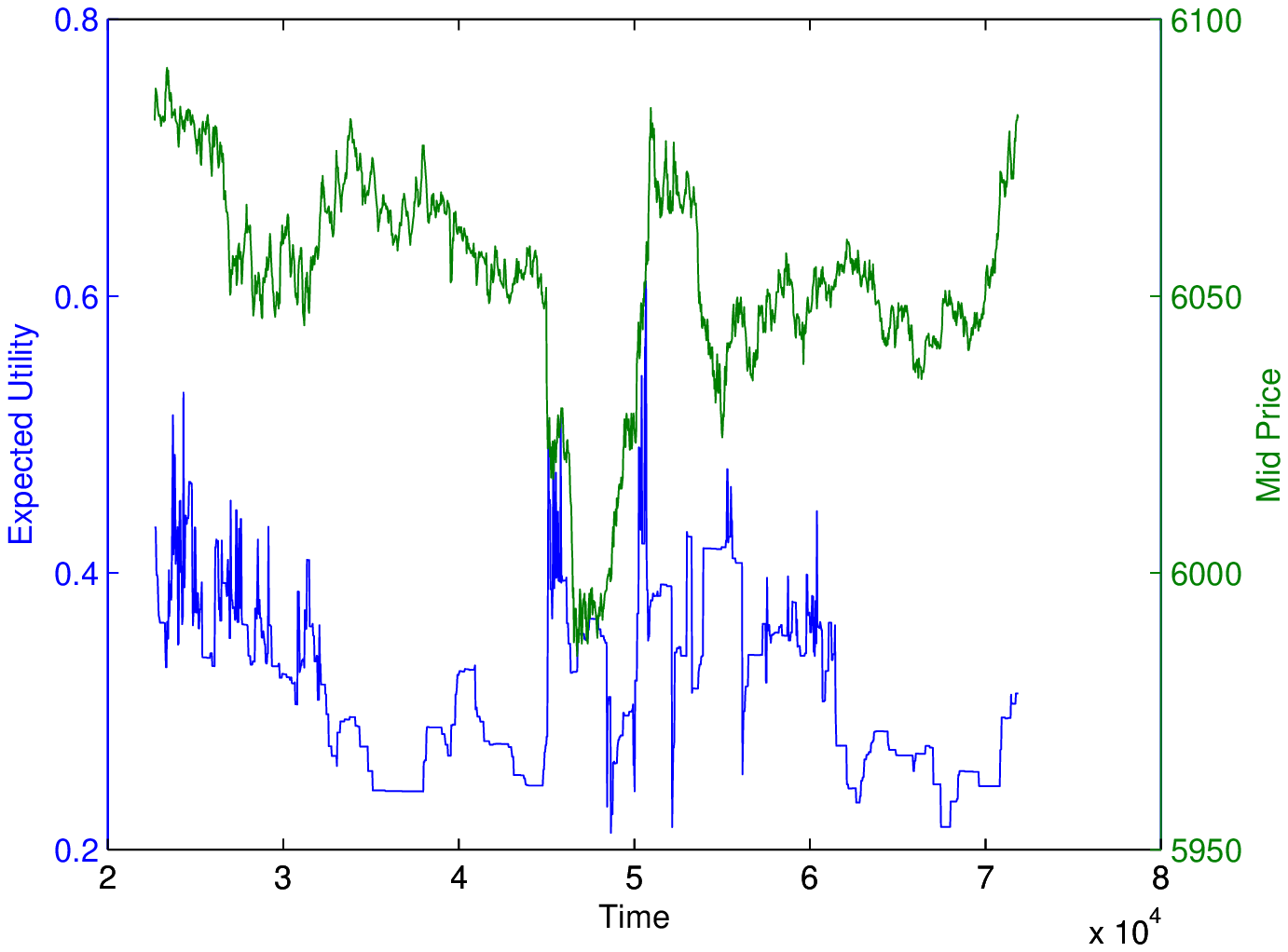}}
\end{figure}
\begin{figure}[H]
  \centering
  \subfloat[GBM Seller]{\includegraphics[scale=0.47]{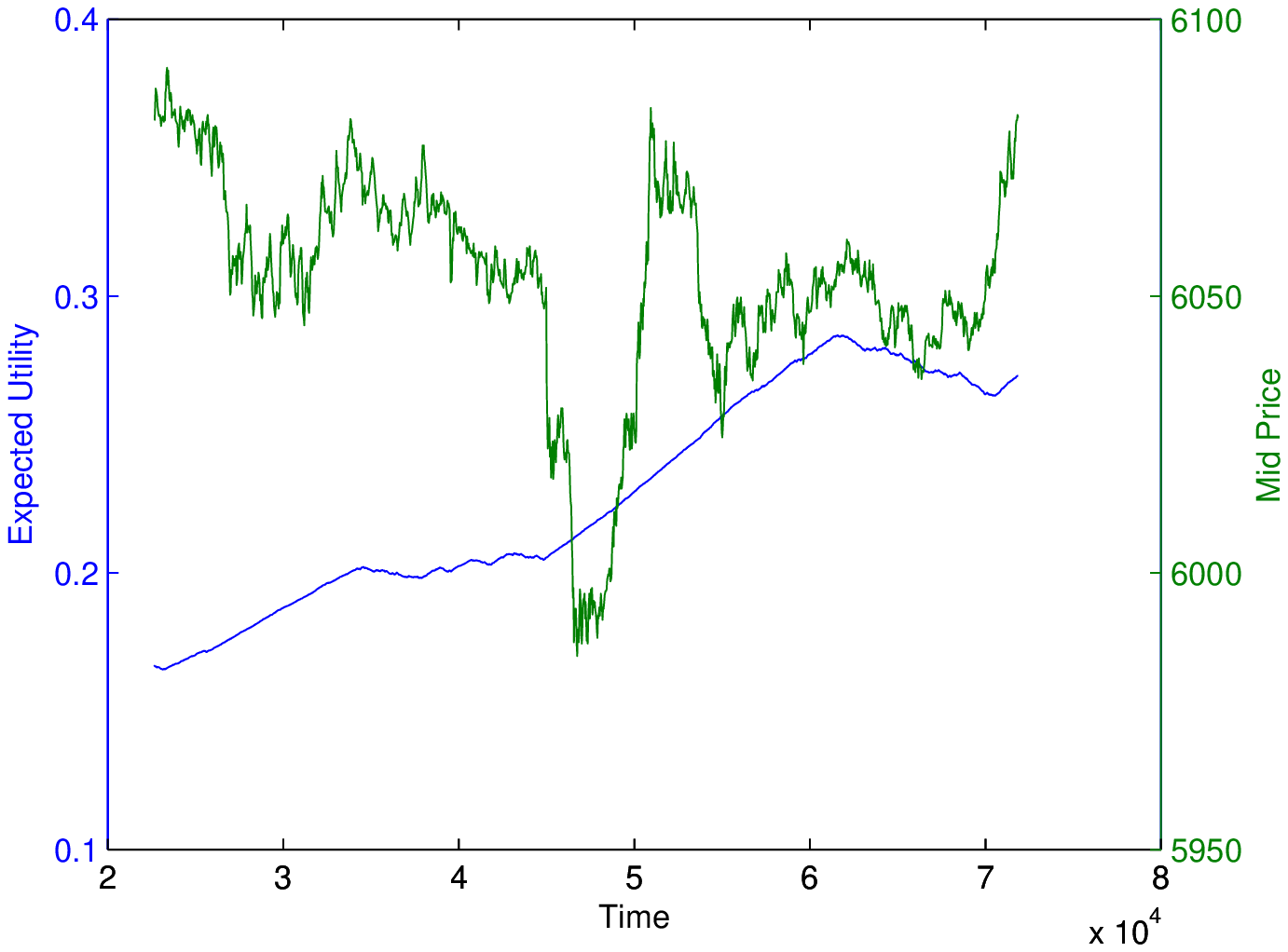}}                
  \subfloat[DEJD Seller]{\includegraphics[scale=0.47]{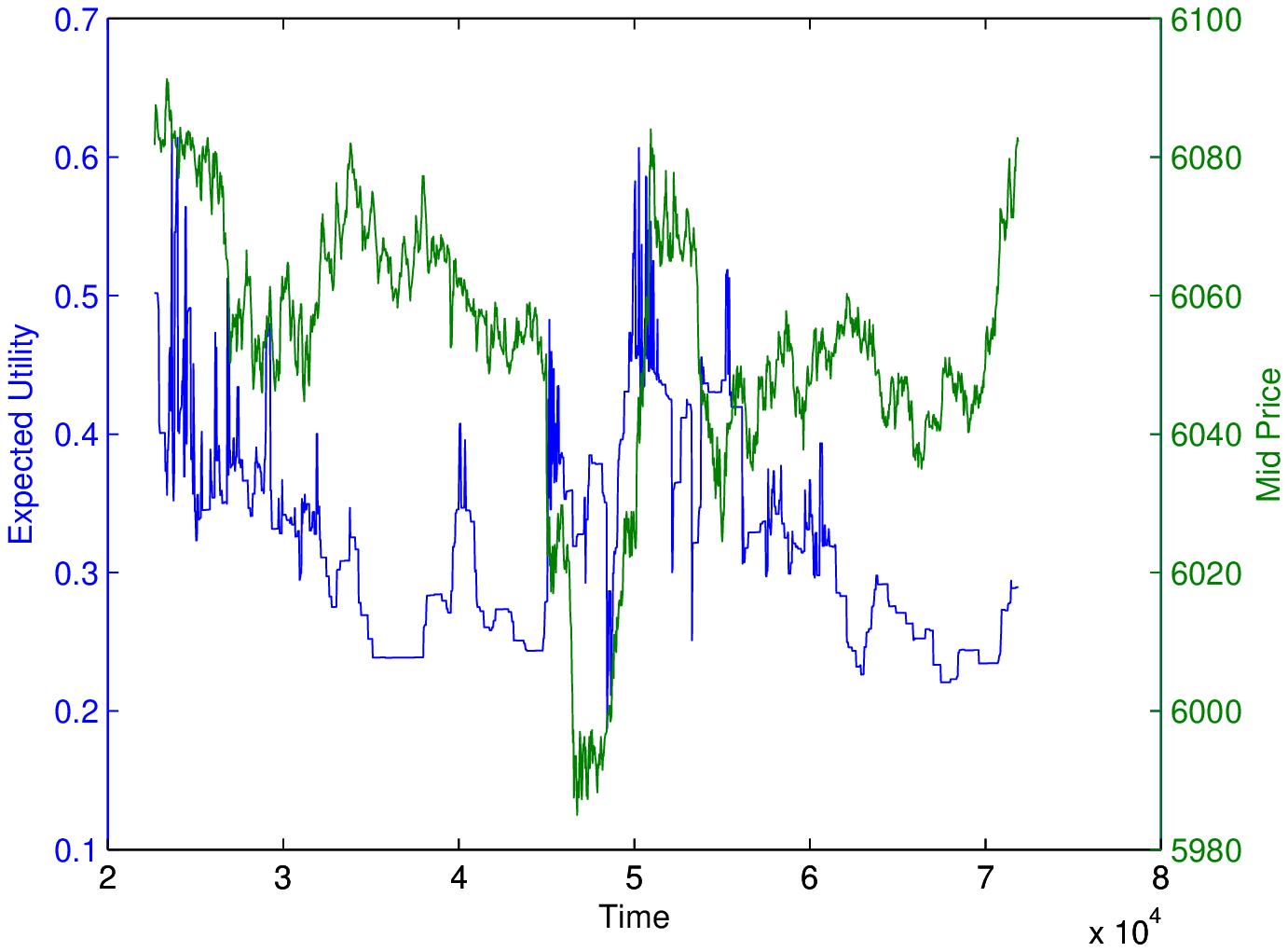}}
  \caption{Expected utilities and mid-price.}
\end{figure}

\section{Proof of Theorem \ref{theorem2} }

A number of preliminary results are due, in order to prove the statement. 
We begin by formulating the It$ \hat{\text{o}} $ lemma for semi-martingales, (eg. 
jump-diffusions), and then apply it to our process to obtain the solution.
\begin{my_lem}[It$ \hat{\text{o}} $'s Lemma for Semi-Martingales]
\footnote{This formulation is essentially the one given in \cite{Protter},  
but has been adapted to the given differential form using the exposition in
\cite{Anatoliy}, as well as \cite{Oksendal}. The convenient notation 
for the derivatives of $ f $ has been inspired by \cite{Shreve}. }
Let $ X $ be a semi martingale, and let $ f $ be real $ C^{2,1} $ function. Then
$ f(X,t) $ is again a semi-martingale, and the following formula holds:
\begin{align}
d f(X_t, t-) & = f_{t}(X_t,t-) \text{d}t+ f_{x}(X_{t-}, t-) \text{d}X_t
+\dfrac{1}{2} f_{xx}(X_{t-}, t-) \text{d}\langle X \rangle^{c}_{t} \nonumber \\
& + \Delta f(X_t, t-)- f_{x}(X_{t-}, t-)\Delta X_t
\intertext{ with the following notations: }
f_{x}(X,t) & := \dfrac{\partial f}{ \partial X }(X,t),
\quad f_{t}(X,t) := \dfrac{\partial f}{ \partial t }(X,t),
\quad f_{xx}(X,t) := \dfrac{\partial^2 f}{ \partial X^2 }(X,t), \nonumber \\
t- & := \lim_{\epsilon \downarrow 0}t-\epsilon,
\quad \Delta X_t := X_t - X_{t-}, \nonumber \\
\quad \Delta f(X_t, t-) & := f(X_t, t-) - f(X_{t-}, t-). \nonumber
\end{align}
Also $ \text{d} \langle X \rangle_t^{c} $ is the differential of the 
quadratic variation process of the continuous part of $ X_t $.
\end{my_lem}
\begin{proof}
A proof is given in \cite{Protter}, p. 78.
\end{proof}
As a consequence the following corollary for the Dol$\acute{\text{e}}$ans-Dade stochastic
 exponential for semi-martingales can be formulated:
\begin{my_cor}[Stochastic Exponential for Semi-martingales]
\footnote{Again, this formulation has been adapted from Theorem 37 in \cite{Protter}, 
p. 84 to be in the more convenient differential form and in the shorter 
form, as given in the theorem's proof also there.}
Let $ X_t $ be a semi-martingale, $X_0 = 0$. Then there exists a unique 
semi-martingale $ Z_t $ satisfying $ \text{d}Z_t =Z_{t-}\text{d} X_t $ with 
$ Z_0 = 1 $. $ Z_t $ is given by:
\begin{equation}
Z_t = \exp\left(X_t - \dfrac{1}{2}\langle X \rangle_t^{c}\right)
\prod_{s\le t}\Big((1+\Delta X_s)\exp(-\Delta X_s)\Big)
\end{equation}
\end{my_cor} 
\begin{proof}
Cf. \cite{Protter}, p.84.
\end{proof}

\begin{proof}[\textbf{Proof of Theorem \ref{theorem2}} ]
Consider the following function
\begin{equation*}
f(x,t):=xe^{\mu t}, \; f_x(x,t)=e^{\mu t}, \; f_{xx}(x,t)=0, 
\; f_t(x,t) = \mu x e^{\mu t} = \mu f(x,t).
\end{equation*}
We look for a solution of the form;
\begin{equation*}
X_t = C_t e^{\mu t}
\end{equation*}
for some semi-martingale $ C_t $. Applying the It$ \hat{\text{o}} $-formula for 
semi-martingales we obtain:
\begin{equation*}
\text{d}X_t  =d f(C_t,t)= e^{\mu t}\text{d}C_t + \mu C_t e^{\mu t}\text{d}t
\end{equation*}
Notice that under the assumption $ X_t = C_t e^{\mu t} $ the SDE can be rewritten as:
\begin{equation*}
\text{d}X_t = C_t e^{\mu t}( \sigma \text{d}B_t + \text{d}J_t + \mu \text{d}t )
\end{equation*}
Comparing the last two equations, we conclude that:
\begin{equation*}
\text{d}C_t = C_t (\sigma \text{d}B_t + \text{d}J_t),
\end{equation*}
and by virtue of the previous corollary, we know that the unique solution for $ C_t $
is given by:
\begin{align*}
C_t & =S\exp\left(\sigma B_t + \sum_{i=1}^{N_t}(V_i-1) - \dfrac{1}{2}\sigma^2t\right)
\prod_{i=1}^{N_t}V_i \exp\big(-(V_i - 1)\big) \\
& = S\exp\left(\sigma B_t - \dfrac{1}{2}\sigma^2 t\right)
\prod_{i=1}^{N_t}V_i,
\intertext{which yields }
X_t & = C_t e^{\mu t} = \exp\left(\left(\mu - \dfrac{\sigma^2}{2}\right)t + \sigma B_t\right)
\prod_{i=1}^{N_t}V_i,
\end{align*}
concluding the proof.\footnote{An intuitive proof of this with omission of technicalities is
given in \cite{Pollard}. }
\end{proof}

Note that \eqref{eq:dejd_process} is the same as:
\begin{equation}
S_t = S \exp\left(\left(\mu - \dfrac{\sigma^2}{2}\right)t + \sigma B_t + \sum_{i=1}^{N(t)}Y_i \right) 
\end{equation}

\section{Proof of Theorem \ref{theorem3} }

\begin{my_def}[Poisson Process]
A right-continuous process $ (N_t)_{t\ge 0} $ with state space $ \mathbb{N}_0 $ is a time homogeneous
 Poisson process with intensity rate $ \lambda $ iff the following is true:
\begin{itemize}
\item[a)] $ N_0=0 $ a.s.;
\item[b)] the process has stationary, independent increments, which are Poisson distributed, i.e.
for all $ s,t\ge 0 $ $ N_{s+t} - N_s \sim \text{Poi(}\lambda t \text{)} $.
\end{itemize}
\end{my_def}

\begin{my_def}[Compound Poisson Process]
\label{cpp}
Let $ (N_t)_{t\ge 0} $ be a time homogeneous Poisson process with intensity rate $ \lambda $
and $ (Y_n)_{n\in\mathbb{N}} $ a family of iid random variables, and let the family further be
independent of $ (N_t)_{t\ge 0} $. The process $ (C_t)_{t\ge 0} $, defined by
\begin{equation*}
C_t := \sum_{i=1}^{N_t} Y_i, \quad t\ge 0
\end{equation*}
is called a compound Poisson process.
\end{my_def}

\begin{my_lem}
For a time homogeneous compound Poisson process $ (C_t)_{t\ge 0 } $ with intensity rate $ \lambda $ and
square-integrable $ Y_1 $ the following holds:
\begin{equation*}
\mathbb{E}[C_t]=\lambda t \mathbb{E}[Y_1] \text{ and } \mathbb{V}[C_t]=\lambda t \mathbb{E}\left[Y_1^2\right].
\end{equation*}
\label{cpp_lemma}
\end{my_lem}

\begin{proof}
Cf. \cite{Meintrup}, Theorem 10.24, iii).
\end{proof}

We will also need a slight variation of this Lemma concerning the moments of a ``Poisson product" of 
iid random variables, which we will prove:
\begin{my_lem}
Let $ (N_t)_{t\ge 0} $ be a time homogeneous Poisson process with intensity rate $ \lambda $ 
and let $ t\ge 0 $ be fixed. Further let $ (Y_i)_{i\in\mathbb{N}} $ be a family of iid random variables,
which is also independent of the Poisson process. Then for the ``Poisson product":
\begin{equation}
P_t := \prod_{i=1}^{N_t} V_i
\label{eq:poisson_prod}
\end{equation}
the following holds:
\begin{align}
\mathbb{E}[P_t] & =\exp\left( t\lambda (\mathbb{E}[V_1]-1) \right)  
\intertext{ and } 
\mathbb{V}[P_t] & =\exp\left( t\lambda (\mathbb{E}[V_1^2]-1)\right) 
-\exp\left( 2 t \lambda (\mathbb{E}[Y_1]-1) \right)
\end{align}
\end{my_lem}

\begin{proof}
We employ the iid property of the family $ (V_i)_{i\in\mathbb{N}} $, 
and its independence of $ N $. We also make use of the density function of a 
Poisson random variable and a representation of the exponential function:
\begin{align*}
\mathbb{E}[P_t] & = \mathbb{E}\left[\prod_{i=1}^{N_t} V_i \right] 
= \sum_{k=0}^{\infty}\mathbb{E}[V_1]^{k}\mathbb{P}(N_t=k)
 = \sum_{k=0}^{\infty}\mathbb{E}[V_1]^{k} \cdot \dfrac{e^{-t\lambda} (t\lambda)^k }{k!}\\
& = e^{-t\lambda} \sum_{k=0}^{\infty} \dfrac{\left(\mathbb{E}[V_1] t\lambda \right)^k}{k!}
= e^{-t\lambda}e^{ \mathbb{E}[V_1] t\lambda} = \exp\left(t\lambda(\mathbb{E}[V_1]-1)\right)
\end{align*}
For the variance we will need the following:
\begin{align*}
\mathbb{E}[P_t^2] & = \mathbb{E}\left[\left(\prod_{i=1}^{N_t} V_i\right)^2 \right] =
\mathbb{E}\left[\prod_{i=1}^{N_t} V_i^2 \right]  =
\sum_{k=0}^{\infty}\mathbb{E}[V_1^2]^{k}\mathbb{P}(N_t=k)\\
& = \exp\left(t\lambda(\mathbb{E}[V_1^2]-1)\right)
\end{align*}
so the variance is
\begin{align*}
\mathbb{V}[P_t] & = \mathbb{E}[P_t^2]-\mathbb{E}[P_t]^2 =
\exp\left(t\lambda(\mathbb{E}[V_1^2]-1)\right) - \exp\left(2t\lambda(\mathbb{E}[V_1]-1)\right)
\end{align*}
\end{proof}

\begin{proof}[\textbf{Proof of Theorem \ref{theorem3} }]
\begin{itemize}
\item[a)] First we show how the first two moments of a double-exponential random variable are
calculated:
\begin{equation*}
\mathbb{E}[Y]=\mathbb{E}\left[ U\xi^{+}-(1-U)\xi^{-}\right] = \dfrac{p}{\eta_1}-\dfrac{q}{\eta_2};
\end{equation*}
For the variance we have:
\begin{align*}
\mathbb{V}[Y] & =\mathbb{E}[(U\xi^{+})^2]-2\mathbb{E}[U(1-U)\xi^{+}\xi^{-}]+\mathbb{E}[((1-U)\xi^{-})^{2}] 
- \mathbb{E}[Y]^2\\
& = \dfrac{2p}{\eta_1^2}+\dfrac{2q}{\eta_2^2} 
- \dfrac{p^2}{\eta_1^2} + 2\dfrac{pq}{\eta_1 \eta_2} - \dfrac{q^2}{\eta_2^2}\\
& = \dfrac{2p}{\eta_1^2}+\dfrac{2q}{\eta_2^2} 
- \dfrac{p(1-q)}{\eta_1^2} + 2\dfrac{pq}{\eta_1 \eta_2} - \dfrac{q(1-p)}{\eta_2^2}\\
& = \left( \dfrac{p}{\eta_1^2} + \dfrac{q}{\eta_2^2}\right)+pq \left( \dfrac{1}{\eta_1}+
\dfrac{1}{\eta_2} \right)^2
\end{align*}
\item[b)] Now we turn to the first two moments of $ V=\exp(Y) $ from Theorem \ref{theorem2}.
\begin{align*}
\mathbb{E}[V] & =\mathbb{E}[\exp(Y)]=\mathbb{E}[\exp(U\xi^{+} - (1-U)\xi^{-})] \\
& = p\mathbb{E}[\exp(\xi^{+})]+q\mathbb{E}[\exp(-\xi^{-})] = p\dfrac{\eta_1}{\eta_1-1} 
+ q \dfrac{\eta_2}{\eta_2+1}
\end{align*}
where in the last step we used the moment generating function of the exponential distribution
(cf. \cite{Meintrup}, p. 102). Note that for the first moment of $ V $ to to be finite,  $ \eta_1 >1$
is required, and for $ \mathbb{E}[V^2] < \infty $, $ \eta_1>2 $ is needed (cf. \cite{Meintrup}, 
p. 103). Note that we have explicitly imposed those constraints in our model.
Now we turn to the variance. First we obtain:
\begin{align*}
\mathbb{E}[V^2] & =\mathbb{E}[\exp(2Y)]=\mathbb{E}[\exp(U2\xi^{+} - (1-U)2\xi^{-})] \\
& = p\mathbb{E}[\exp(2\xi^{+})]+q\mathbb{E}[\exp(-2\xi^{-})] = p\dfrac{\eta_1}{\eta_1-2} 
+ q \dfrac{\eta_2}{\eta_2+2},
\end{align*}
and again, the last step was produced by using the moment generating function 
of an exponential distribution. So, in summary, the variance of $ V $ is given by:
\begin{align*}
\mathbb{V}[V] & =\mathbb{E}[V^2] - \mathbb{E}[V]^2 \\
& = p\dfrac{\eta_1}{\eta_1-2} + q \dfrac{\eta_2}{\eta_2+2} - 
\left( p\dfrac{\eta_1}{\eta_1-1} + q \dfrac{\eta_2}{\eta_2+1} \right)^2
\end{align*}
\item[c)] This has been considered in the previous lemma.
\item[d)] Next we turn to the moments of the stochastic process itself:
\begin{align*}
\mathbb{E}[S_t] & =\mathbb{E}\left[S\exp\left(\left(\mu-\dfrac{\sigma^2}{2}\right)t 
+ \sigma B_t\right)\prod_{i=^1}^{N(t)}V_i \right] \\
& =S\mathbb{E}\left[\exp\left(\left(\mu-\dfrac{\sigma^2}{2}\right)t 
+ \sigma B_t\right)\right] \mathbb{E}\left[\prod_{i=^1}^{N(t)}V_i\right] \\
& = S \exp(\mu t)\mathbb{E}[P_t] 
\end{align*}
where $ P_t $ is the ``Poisson product" from \eqref{eq:poisson_prod}, and the term $ e^{\mu t} $
was obtained as the first moment of a log-normally distributed random variable,
which corresponds to distribution of the GBM process in time. 
Next, we look at the second moment of the DEJD:
\begin{align*}
\mathbb{E}[S_t^2] & = \mathbb{E}\left[S^2\exp\left(2\left(\mu-\dfrac{\sigma^2}{2}\right)t 
+ 2\sigma B_t\right)\prod_{i=^1}^{N(t)}V_i^2 \right]\\
& = S^2\exp(2\mu t + \sigma^2 t ) \mathbb{E}[P_t^2],
\end{align*}
so in summary the variance is given by:
\begin{align*}
\mathbb{V}[S_t]& = \mathbb{E}[S_t^2]-\mathbb{E}[S_t]^2
 = S^2 e^{2\mu t}\left(e^{\sigma^2 t}\mathbb{E}[P_t^2] - \mathbb{E}[P_t]^2\right)
\end{align*}
\item[e)] Finally, after characterizing the process itself,
we take a look at the log-returns $ l(\Delta t) $, which are of particular importance
for our empirical work, as we calibrate our model parameters, so as to match the
rolling volatility and mean of the observed log-returns in the market:
\begin{align*}
\mathbb{E}[l(\Delta)] & = \mathbb{E}\left[\log\left(
\dfrac{S\exp\left(\left(\mu-\dfrac{\sigma^2}{2}\right)(t+\Delta t) 
+ \sigma B_{t+\Delta t} + \sum_{i=1}^{N_{t+\Delta t}}Y_i\right)}
{S\exp\left(\left(\mu-\dfrac{\sigma^2}{2}\right)t) 
+ \sigma B_{t} + \sum_{i=1}^{N_{t}}Y_i\right)}\right)
 \right]\\
& = \mathbb{E}\left[\left(\mu-\dfrac{\sigma^2}{2}\right)\Delta t
+ \sigma B_{\Delta t} + \sum_{i=1}^{N_{\Delta t}}Y_i \right]\\
& = \left(\mu -\dfrac{\sigma^2}{2}\right)\Delta t + \lambda \Delta t \mathbb{E}[Y] \\
& = \left(\mu -\dfrac{\sigma^2}{2}\right)\Delta t + \lambda \Delta t 
\left(\dfrac{p}{\eta_1} - \dfrac{q}{\eta_2}\right)
\end{align*}
where the penultimate step was achieved by observing that $ \sum_{i=1}^{N_{\Delta t}} Y_i $
is a compound Poisson process and together with Lemma \ref{cpp_lemma}.

We obtain the variance as the sum of two independent random variables, namely a
 scaled Brownian motion with drift and a compound Poisson process:
\begin{align*}
\mathbb{V}\left[l(\Delta t) \right] & =
\mathbb{V}\left[\left(\mu-\dfrac{\sigma^2}{2}\right)\Delta t
+ \sigma B_{\Delta t} + \sum_{i=1}^{N_{\Delta t}}Y_i \right]\\
& = \mathbb{V}\left[\left(\mu-\dfrac{\sigma^2}{2}\right)\Delta t
+ \sigma B_{\Delta t}\right] + \mathbb{V}\left[\sum_{i=1}^{N_{\Delta t}}Y_i \right]\\
& = \sigma^2 \Delta t + \lambda \Delta t \left(\dfrac{2p}{\eta_1^2} + 
\dfrac{2q}{\eta_2^2} \right)
\end{align*}

\end{itemize}

\end{proof}

\begin{landscape}

\section{Descriptive Statistics}

\begin{figure}[!ht]
\begin{center}
\begin{scriptsize}

\begin{tabular}{ c | c c c c | c c c c || c c c c | c c c c }
\hline
 &
\multicolumn{8}{c||}{GBM} &
\multicolumn{8}{c}{DEJD}\\
\hline
 & \multicolumn{4}{c|}{Bid} &
\multicolumn{4}{c||}{Ask} &
\multicolumn{4}{c|}{Bid} &
\multicolumn{4}{c}{Ask} \\
\hline
Date & $\overline{r}$ & $\sigma_r$ & $\overline{U}$ & $\sigma_U$ & $\overline{r}$ & $\sigma_r$ & $\overline{U}$ & $\sigma_U$ & $\overline{r}$ & $\sigma_r$ & $\overline{U}$ & $\sigma_U$ & $\overline{r}$ & $\sigma_r$ & $\overline{U}$ & $\sigma_U$ \\
\hline
6/7/2010&1.6697&1.5638&0.3343&0.0534&1.2232&0.961&0.3791&0.0348&2.5968&0.4754&0.3106&0.0668&2.7041&0.4278&0.3114&0.0742\\
6/8/2010&1.7938&1.5542&0.3516&0.0011&1.9258&1.8318&0.3742&0.0197&2.5922&1.6858&0.3807&0.0553&2.4968&1.6743&0.3953&0.0573\\
6/9/2010&1.4389&1.2&0.3492&0.0012&1.4433&1.0755&0.3534&0.0149&0.4052&0.0078&0.2457&0.0134&0.3832&0.0047&0.217&0.0069\\
6/10/2010&0.8532&0.8239&0.3963&0.0462&1.8643&1.7271&0.3165&0.0465&2.0466&0.9349&0.3361&0.0583&2.1711&1.0722&0.3354&0.0534\\
6/11/2010&1.8339&1.6336&0.3107&0.0015&4.2609&4.8068&0.2276&0.0378&2.3427&0.9756&0.3203&0.0615&2.3329&0.8856&0.3259&0.0737\\
6/14/2010&6.6012&7.2139&0.1279&0.0013&1.2499&0.9366&0.3096&0.0219&0.3846&0.0061&0.2303&0.0102&0.3762&0.0063&0.2266&0.0113\\
6/15/2010&1.052&0.8965&0.3239&0.0371&2.3865&2.6579&0.2521&0.0176&0.4021&0.0025&0.2508&0.0042&0.4021&0.0025&0.2509&0.0041\\
6/16/2010&11.5236&14.185&0.072&0.0009&1.3205&1.025&0.2986&0.0145&1.6776&0.426&0.3123&0.0589&2.1813&0.7166&0.2892&0.0686\\
6/17/2010&4.262&5.6016&0.1753&0.0012&1.5435&1.6336&0.3001&0.0153&1.7524&1.1442&0.3314&0.0675&1.9127&1.3945&0.3243&0.0646\\
6/18/2010&22.2285&28.0068&0.0175&0.0093&8.8825&13.3641&0.1291&0.1044&1.5505&0.6049&0.3945&0.0918&1.1383&0.4843&0.44&0.1014\\
6/22/2010&1.4545&0.8066&0.345&0.0002&13.7943&13.4171&0.0854&0.0032&1.3359&0.4592&0.3831&0.0483&1.3734&0.4972&0.3819&0.0419\\
6/23/2010&1.3237&1.7857&0.3549&0.0451&2.9182&3.4364&0.2497&0.0179&0.4014&0.0132&0.2477&0.0222&0.4068&0.014&0.265&0.025\\
6/24/2010&16.0602&26.9084&0.101&0.0736&13.8511&19.5797&0.0912&0.0305&0.3802&0.0095&0.231&0.0161&0.3785&0.0101&0.2302&0.019\\
6/25/2010&4.6223&4.7647&0.211&0.0016&3.4199&3.4273&0.2549&0.0348&2.857&0.7843&0.2998&0.0671&2.5293&0.5724&0.3112&0.0625\\
6/28/2010&1.7127&1.5893&0.2786&0.0007&3.7916&3.8808&0.1966&0.0235&2.1239&0.44&0.2915&0.0527&2.0837&0.4705&0.294&0.0548\\
6/29/2010&1.2483&2.1697&0.3852&0.0123&1.4386&1.9251&0.356&0.0263&2.5392&1.9022&0.3322&0.0867&2.6234&1.9291&0.3251&0.0742\\
6/30/2010&26.0247&50.6983&0.0548&0.0136&3.0267&5.1741&0.2782&0.0227&2.5383&2.0542&0.3331&0.1372&2.5013&2.0531&0.3295&0.0993\\
7/1/2010&2.1873&2.3846&0.31&0.001&1.4374&1.1859&0.3586&0.0159&0.4077&0.006&0.2618&0.0121&0.3763&0.0037&0.2151&0.0056\\
7/2/2010&6.3956&17.3426&0.2048&0.0017&3.862&9.8311&0.2759&0.0259&3.2363&2.5481&0.3152&0.1132&3.1933&2.6095&0.3152&0.0985\\
7/5/2010&1.1382&1.5598&0.17&0.002&10.2659&19.6048&0.0237&0.0072&0.395&0.0159&0.2358&0.0368&0.3885&0.011&0.2233&0.0234\\
7/6/2010&34.3653&43.3035&0.0191&0.0013&6.8719&7.5546&0.1344&0.0144&1.7868&0.7025&0.3126&0.0542&1.7482&0.6176&0.3141&0.0554\\
7/7/2010&1.965&1.3795&0.2742&0.0217&6.6386&7.1599&0.1634&0.0323&2.2267&0.4787&0.3021&0.0498&2.5558&0.5673&0.2914&0.0527\\
7/8/2010&2.0677&1.8456&0.222&0.0005&7.5357&8.391&0.1011&0.017&1.4505&0.3396&0.3087&0.0523&1.5376&0.4038&0.2987&0.0517\\
7/9/2010&0.703&0.4733&0.3216&0.0046&1.8586&1.4371&0.2203&0.0399&1.4362&0.2944&0.2874&0.0516&1.3547&0.2469&0.2978&0.0487\\
7/12/2010&6.0655&7.8103&0.1004&0.0502&1.1233&0.8429&0.2722&0.0285&1.505&0.32&0.2861&0.0567&1.4453&0.2839&0.2841&0.0543\\
\hline
\end{tabular}
\end{scriptsize}
\end{center}
\caption{Descriptive statistics per day: $ \overline{r} $ is the mean impatience rate, 
$ \sigma_r $ is its standard deviation, $ \overline{U} $ is the expected utility and
$ \sigma_U $ is its standard deviation. Notice that $ \sigma_r $ is much more lower for the DEJD.}
\label{fig:d_stats_1}
\end{figure}

\begin{figure}[!ht]
\begin{center}
\begin{scriptsize}

\begin{tabular}{ c | c c c c | c c c c || c c c c | c c c c }
\hline
 &
\multicolumn{8}{c||}{GBM} &
\multicolumn{8}{c}{DEJD}\\
\hline
 & \multicolumn{4}{c|}{Bid} &
\multicolumn{4}{c||}{Ask} &
\multicolumn{4}{c|}{Bid} &
\multicolumn{4}{c}{Ask} \\
\hline
Date & $\overline{r}$ & $\sigma_r$ & $\overline{U}$ & $\sigma_U$ & $\overline{r}$ & $\sigma_r$ & $\overline{U}$ & $\sigma_U$ & $\overline{r}$ & $\sigma_r$ & $\overline{U}$ & $\sigma_U$ & $\overline{r}$ & $\sigma_r$ & $\overline{U}$ & $\sigma_U$ \\
\hline
7/13/2010&2.5726&4.5379&0.227&0.0393&11.6462&30.8123&0.0989&0.0173&1.3068&0.7884&0.3331&0.0811&1.4829&0.9567&0.3193&0.0671\\
7/14/2010&66.3458&83.8717&0.0032&0.002&2.6034&2.9797&0.2073&0.0333&0.3939&0.013&0.2296&0.0244&0.3958&0.0115&0.248&0.02\\
7/15/2010&1.7529&1.8065&0.2833&0.0445&24.0865&35.2304&0.0425&0.0097&0.3987&0.0107&0.2469&0.0158&0.3987&0.0107&0.247&0.0157\\
7/16/2010&8.1558&13.0939&0.1375&0.0016&1.5682&1.619&0.3109&0.0165&0.413&0.0138&0.2527&0.0276&0.4033&0.0123&0.2473&0.0202\\
7/19/2010&1.6708&2.1856&0.2986&0.0826&2.4383&2.9992&0.2468&0.0384&0.418&0.0337&0.2371&0.0263&0.384&0.013&0.2203&0.024\\
7/21/2010&5.5933&10.7429&0.2368&0.1321&7.1283&8.1209&0.148&0.027&0.4117&0.0143&0.2535&0.0254&0.3862&0.0101&0.2228&0.0161\\
7/22/2010&0.7802&0.666&0.3541&0.0206&4.61&6.0582&0.1822&0.0246&0.3815&0.0089&0.227&0.016&0.383&0.0089&0.2312&0.0168\\
7/23/2010&6.9123&8.6572&0.1398&0.0012&3.4125&4.049&0.2419&0.0248&2.2171&0.7722&0.3184&0.0929&2.3033&0.8736&0.3115&0.0798\\
7/26/2010&0.6593&0.642&0.3751&0.0202&4.4491&5.443&0.1689&0.021&2.1006&0.4758&0.2792&0.0697&1.9081&0.4389&0.2903&0.0679\\
7/27/2010&7.7934&13.9949&0.1185&0.0841&6.6015&16.2536&0.174&0.102&0.3842&0.0092&0.229&0.0166&0.381&0.0075&0.2289&0.0123\\
7/28/2010&8.2239&9.3875&0.0901&0.0008&2.3513&2.2435&0.2132&0.0112&1.8266&0.3681&0.2875&0.0544&1.8567&0.383&0.2822&0.0519\\
7/29/2010&15.1407&22.151&0.0632&0.0007&2.551&2.6095&0.2452&0.017&1.6739&0.8145&0.3183&0.0613&1.8035&1.0716&0.3153&0.0592\\
8/3/2010&4.4311&6.3508&0.1277&0.0006&14.0899&30.2006&0.1576&0.1154&1.9171&0.843&0.2684&0.0686&1.916&0.7999&0.2687&0.0693\\
8/4/2010&23.3697&41.5286&0.0327&0.0004&28.1208&57.8002&0.0324&0.0062&1.5608&0.8342&0.3474&0.133&1.4836&0.7579&0.3518&0.1102\\
8/5/2010&7.3107&10.2559&0.0777&0.0194&1.9764&2.0087&0.2044&0.0081&0.4424&0.1187&0.2247&0.0164&0.3816&0.0123&0.2073&0.0484\\
8/6/2010&3.5688&8.4808&0.2822&0.1811&5.8963&23.0231&0.2299&0.0527&0.4194&0.0037&0.2666&0.0097&0.4194&0.0037&0.2666&0.0097\\
8/9/2010&0.8401&0.8582&0.261&0.0069&0.9572&1.1474&0.2637&0.0604&1.1236&0.225&0.2818&0.0529&1.1571&0.2477&0.2807&0.0578\\
8/10/2010&4.3673&10.5804&0.1972&0.0593&3.3188&5.7592&0.2232&0.0387&2.0107&0.7966&0.2951&0.0722&2.0289&0.7885&0.2953&0.0699\\
8/11/2010&1.3226&1.9279&0.3206&0.1137&1.2163&0.961&0.2964&0.0256&0.4032&0.0066&0.2551&0.0143&0.4155&0.0074&0.2665&0.0143\\
8/12/2010&1.7891&2.0906&0.2748&0.0346&3.9149&6.4365&0.2044&0.0273&0.4017&0.0083&0.2489&0.0151&0.398&0.0098&0.2452&0.0201\\
8/16/2010&1.0649&0.9926&0.3472&0.016&2.6688&3.0557&0.2455&0.0239&0.3825&0.008&0.236&0.014&0.3918&0.0077&0.2337&0.0126\\
8/17/2010&15.4901&23.2717&0.0447&0.013&2.0762&2.1442&0.2349&0.0961&0.3995&0.0062&0.2381&0.0115&0.3785&0.0032&0.2129&0.0051\\
8/18/2010&2.1794&2.4913&0.2218&0.0113&9.6789&13.3676&0.0875&0.0139&0.4078&0&0.2606&0&0.4079&0&0.2607&0.0001\\
8/19/2010&11.8686&23.3092&0.1226&0.0011&44.7998&88.3008&0.0664&0.084&2.9749&2.7303&0.3225&0.0813&3.0228&2.6258&0.3226&0.0794\\
8/20/2010&1.8464&1.7797&0.2703&0.0414&1.4503&1.5248&0.3003&0.0312&2.1296&0.8265&0.2908&0.0713&2.4249&0.8283&0.2785&0.0801\\
\hline
\hline
Mean&7.3128&10.6631&0.2144&0.0262&6.0309&9.8203&0.2166&0.0319&1.3374&0.5272&0.2858&0.0483&1.3501&0.537&0.2829&0.0462\\
\end{tabular}
\end{scriptsize}
\end{center}
\caption{Continued - descriptive statistics per day: $ \overline{r} $ is the mean 
impatience rate, 
$ \sigma_r $ is its standard deviation, $ \overline{U} $ is the expected utility and
$ \sigma_U $ is its standard deviation. Notice that $ \sigma_r $ is much more lower for the DEJD.}
\label{fig:d_stats_2}
\end{figure}

\end{landscape}

\section{Limit Order Book}

\begin{figure}[!ht]
\begin{center}
\textbf{The limit order book at a point in time}

\setlength{\unitlength}{.8cm}
\begin{picture}(16,16)
\put(1.5,14){$\text{Ask sizes}$}
\put(1,7){\vector(0,1){7}}	
\put(1,7){\vector(0,-1){7}}	
\put(1.5,0){$\text{Bid sizes}$}
\put(1,7){\vector(1,0){15}}	
\put(2,7){\line(0,-1){6}}	
\put(2,1){\line(1,0){1}}	
\put(3,7){\line(0,-1){6}}	

\put(4,7){\line(0,-1){2}}	
\put(4,5){\line(1,0){1}}	
\put(5,7){\line(0,-1){2}}	

\put(6,7){\line(0,-1){3}}	
\put(6,4){\line(1,0){1}}	
\put(7,7){\line(0,-1){3}}	

\put(8,7){\line(0,1){2}}	
\put(8,9){\line(1,0){1}}	
\put(9,7){\line(0,1){2}}	

\put(10,7){\line(0,1){2}}	
\put(10,9){\line(1,0){1}}	
\put(11,7){\line(0,1){2}}	

\put(12,7){\line(0,1){4}}	
\put(12,11){\line(1,0){1}}	
\put(13,7){\line(0,1){4}}	

\put(2.5,6.9){\line(0,1){.2}}
\put(2.25,7.35){$\text{2.5}$}

\put(4.5,6.9){\line(0,1){.2}}
\put(4.25,7.35){$\text{3.00}$}

\put(6.5,6.9){\line(0,1){.2}}
\put(6.25,7.35){$\text{3.50}$}

\put(7.5,10){\vector(0,-1){2}}
\put(6.75,10.5){$ \text{Mid price} $}
\put(7.5,6.8){\line(0,1){.4}}

\put(8.5,6.9){\line(0,1){.2}}
\put(8.25,6.25){$\text{4.00}$}

\put(10.5,6.9){\line(0,1){.2}}
\put(10.25,6.25){$\text{4.50}$}

\put(12.5,6.9){\line(0,1){.2}}
\put(12.25,6.25){$\text{5.00}$}

\put(14,7.35){$ \text{Asset price} $}

\put(.9,1){\line(1,0){.2}}
\put(.9,2){\line(1,0){.2}}
\put(.9,3){\line(1,0){.2}}
\put(.9,4){\line(1,0){.2}}
\put(.9,5){\line(1,0){.2}}
\put(.9,6){\line(1,0){.2}}
\put(.9,7){\line(1,0){.2}}
\put(.9,8){\line(1,0){.2}}
\put(.9,9){\line(1,0){.2}}
\put(.9,10){\line(1,0){.2}}
\put(.9,11){\line(1,0){.2}}
\put(.9,11){\line(1,0){.2}}
\put(.9,12){\line(1,0){.2}}
\put(.9,13){\line(1,0){.2}}

\put(.575,.75){$ \text{6} $}
\put(.575,1.75){$ \text{5} $}
\put(.575,2.75){$ \text{4} $}
\put(.575,3.75){$ \text{3} $}
\put(.575,4.75){$ \text{2} $}
\put(.575,5.75){$ \text{1} $}
\put(.575,6.75){$ \text{0} $}
\put(.575,7.75){$ \text{1} $}
\put(.575,8.75){$ \text{2} $}
\put(.575,9.75){$ \text{3} $}
\put(.575,10.75){$ \text{4} $}
\put(.575,11.75){$ \text{5} $}
\put(.575,12.75){$ \text{6} $}

\end{picture}
\end{center}

\caption{A stylized snapshot of the LOB with three levels of information at an arbitrary point in time. 
The mid-price is 3.75, and is the mid-point between the highest bid and 
the lowest ask offer, which are in this case 3.50 and 4.00 respectively. 
The block sizes represent the total number of orders at each level in the book.
Block sizes below the abscissa represent bid-order sizes, while above - ask-order sizes.}
\label{fig:theLOB}
\end{figure}
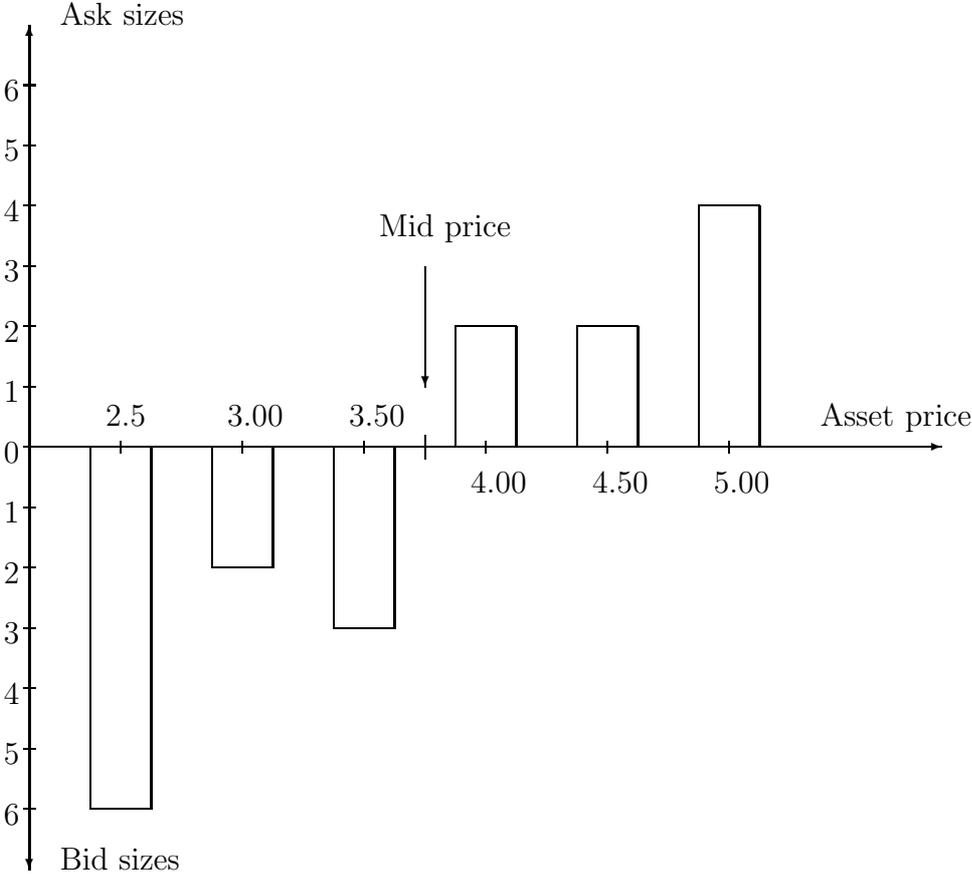
The limit order book (LOB) is the collection of buy and sell orders at any point in time. 
We will explain how it works from the perspective of a buyer. Consider the example of a 
stylized LOB given in Figure \ref{fig:theLOB}. There are six buy orders at a price of 2.50, two 
buy orders at a price of 3.00, and three orders to buy at 3.50. Note that we do not know 
if at each level, the order sizes represent order submission by a single participant, or are simply aggregated 
by the exchange by their limit price. For our purposes it is not important who exactly places which 
orders at which level, so we can safely assume that all orders at each level are placed by 
a single trader. Currently the best buy order is the one at 3.50 for three shares. 
Assume a new buyer wants to enter the market. They could place a limit order at 3.50 or less, 
but will have to wait before the current orders at 3.50 are matched by a seller, or are cancelled by the
respective buyer. They could alternatively place a limit order between 3.50 and 4.00, thus narrowing 
the bid-ask spread, but declaring that they are willing to pay more than the current best bid
 offer. This would move the indicative mid-price up. Or, lastly, they could place a market order.
 Assume they place a market order for four shares. In this scenario the two limit sell orders 
 at four, and the two limit sell orders at 4.50 will be executed, and the order be fully filled. 
 The outermost ask level at 5.00 will become the best ask offer, and the mid-price will move 
 from 3.75 to 4.25. At last, consider the following strategy for our hypothetical buyer, who
 wants to buy four shares at a price of four. They could place a market order for two shares, 
 which will immediately be matched by the best ask price at 4.00, and also place a limit buy 
 order at 4.00 for another two shares, thus becoming the best bidder at 4.00. 
 Notice also
 that in the last scenario, the limit buy order at 2.50 will become invisible to a market 
 participant who only has access to the best three offers on each side of the mid-price. In 
 exchange, potentially a new limit sell order at above 5.00 will be illuminated as the best 
 sell order got filled and the participant is entitled to only see the best three sell orders.

\section{Empirical Fitting}
In this section we first give the reader an idea of what the raw LOB data looks like,
what particular type of data we have had at our avail, and what is an efficient way to 
organize it for further analysis. Next, we will present our method of calibration of 
the model parameters to market data and finally we describe the specificities of 
our fitting procedure, with which the impatience rate is estimated.

An excerpt of raw LOB data for August 20, 2010 is given in Figure \ref{fig:raw_LOB}:

\begin{figure}[!ht]
\begin{center}
\begin{tabular}{ c c l l l l}
\hline
Time (ms) & Level & Bid Limit & Ask Limit & Bid size & Ask size \\
\hline
$ \cdots $ & $ \cdots $ & $ \cdots $ & $ \cdots $ & $ \cdots $ & $ \cdots $ \\
13:59:32:367; & 3; & 6022; & 6025.5; & 25; & 32 \\
13:59:32:367; & 4; & 6021.5; & 6026; & 34; & 43 \\
13:59:32:367; & 0; & 6023.5; & 6024; & 4; & 2 \\
13:59:32:383; & 1; & 6023; & 6024.5; & 14; & 12 \\
13:59:32:383; & 2; & 6022.5; & 6025; & 14; & 23 \\
13:59:32:383; & 3; & 6022; & 6025.5; & 25; & 30 \\
13:59:32:397; & 4; & 6021.5; & 6026; & 34; & 21 \\
13:59:32:397; & 0; & 6024; & 6024; & 1; & 2 \\
13:59:32:413; & 1; & 6023.5; & 6024.5; & 8; & 12 \\
13:59:32:413; & 2; & 6023; & 6025; & 15; & 23 \\
13:59:32:413; & 3; & 6022.5; & 6025.5; & 15; & 30 \\
13:59:32:430; & 4; & 6022; & 6026; & 25; & 21 \\
13:59:32:430; & 0; & 6024; & 6024.5; & 1; & 14 \\
13:59:32:430; & 1; & 6023.5; & 6025; & 8; & 23 \\
13:59:32:430; & 2; & 6023; & 6025.5; & 15; & 30 \\
13:59:32:430; & 3; & 6022.5; & 6026; & 15; & 21 \\
13:59:32:443; & 4; & 6022; & 6026.5; & 25; & 15 \\
13:59:32:443; & 3; & 6022.5; & 6026; & 16; & 21 \\
13:59:32:460; & 4; & 6022; & 6026.5; & 26; & 15 \\
13:59:32:460; & 0; & 6024; & 6024.5; & 1; & 13 \\
13:59:32:460; & 1; & 6023.5; & 6025; & 8; & 22 \\
13:59:32:460; & 3; & 6022.5; & 6026; & 16; & 22 \\
13:59:32:460; & 4; & 6022; & 6026.5; & 26; & 16 \\
13:59:32:477; & 0; & 6024; & 6024.5; & 2; & 13 \\
13:59:32:477; & 1; & 6023.5; & 6025; & 10; & 22 \\
13:59:32:477; & 4; & 6022; & 6026.5; & 32; & 16 \\
13:59:32:490; & 0; & 6024; & 6024.5; & 2; & 11 \\
13:59:32:490; & 2; & 6023; & 6025.5; & 15; & 28 \\
13:59:32:490; & 3; & 6022.5; & 6026; & 16; & 17 \\
13:59:32:490; & 4; & 6022; & 6026.5; & 32; & 17 \\
13:59:32:490; & 0; & 6024; & 6024.5; & 3; & 11 \\
$ \cdots $ & $ \cdots $ & $ \cdots $ & $ \cdots $ & $ \cdots $ & $ \cdots $ \\
\hline
\end{tabular}
\end{center}
\caption{Raw data of the nearest DAX future (FDAX) of August 20, 2010 with a depth of six 
levels - zero to five - on each side of the book.}
\label{fig:raw_LOB}
\end{figure}

The data consists of a time stamp in milliseconds, the level at which a change in the LOB
has occurred, followed by the updated bid limit price, ask limit price, bid order size and
ask order size. Notice that the data simply gives an update at each level if something changes 
and is in itself not the LOB. Much rather it is our task to reconstruct from this data set the 
actual snapshot of the LOB at each point in time. In the example of Figure \ref{fig:raw_LOB}, 
the first three orders timestamped 13:59:32:367 update the order at levels
three, four and zero. Notice that level zero is designated to be the level at which the best bid and the 
best ask orders are given. Thus, the spread at the beginning of this excerpt is 0.5 - the 
difference between 6023.5 and 6024 - which is also the tick-size, i.e. the minimal price 
increment, of the DAX future. Then, at time 13:59:32:383 updates to the levels one and 
two are seen (previous values not given in this excerpt), and an update to the ask size in level 
three - it has gone down from 32 to 30.
Of particular interest is the event of order execution. In our example we see an order being 
executed at level zero at 13:59:32:397, which is indicated by a price matching of, in this case,
6024, which was the previous best ask offer. The transaction size is the lesser of the 
two order sizes at that level, which is one in the given example. We don't know if it is 
the agent with the previous best bid offer of 6023.5 who has cancelled all or part of 
their orders and placed a market order, or a limit buy order at 6024, which was immediately 
executed. What would seem more plausible in this case is that a new agent has come to 
the market, placing a market order for one futures contract, as we can see in the 
follow-up update at 13:59:32:413 that there are now even more orders at 6023.5, which 
are now on level one (previously zero) on the bid side, i.e. this price is no longer the 
best bid offer. Notice also how in the time immediately after the order execution, 
an automatic update of levels takes place on the ask side - what was previously at level three, is now at level two,
what was previously at level four, is now at level three and so forth. 
It is unfortunate that our data does not contain order flags, which explain exactly what has 
happened in each time-step, i.e. containing explicit information whether an order has been 
executed, or more interestingly since it cannot be reliably reconstructed from this data, if
an order has been cancelled. However, using interpretation techniques as demonstrated in
the previous paragraph, one can reconstruct the LOB at each time step and have a 
complete picture of what the LOB looks like at each point in time. This is also 
the way we organize our data for further analysis. In order to have a better 
understanding of how exactly we have reconstructed the LOB from the raw data, 
please see Appendix C, where the data from Figure \ref{fig:raw_LOB} has been 
composed in a more convenient way for further analysis.

\begin{landscape}
\section{Formatted LOB Snapshots}
\begin{figure}[!ht]

\begin{center}


\begin{scriptsize}

\begin{tabular}{c | r r p{.375cm} p{.375cm} | r r p{.375cm} p{.375cm}
 | r r p{.375cm} p{.375cm} | r r p{.375cm} p{.375cm}| r r p{.375cm} p{.375cm}}
\hline
\hline
 & \multicolumn{4}{c}{Level 0} &
\multicolumn{4}{|c}{Level 1} &
\multicolumn{4}{|c}{Level 2} &
\multicolumn{4}{|c}{Level 3} &
\multicolumn{4}{|c}{Level 4} \\
\hline
T & BP & AP & BS & AS
 & BP & AP & BS & AS
 & BP & AP & BS & AS
 & BP & AP & BS & AS
 & BP & AP & BS & AS \\
\hline		 
13:59:32:367 & 6023.5 & 6024 & 4 & 2 
			& 0 & 0 & 0 & 0 
			& 0 & 0 & 0 & 0 
			& 6022 & 6025.5 & 25 & 32 
			& 6021.5 & 6026 & 34 & 43 \\
13:59:32:383 & 6023.5 & 6024 & 4 & 2 
			& 6023 & 6024.5 & 14 & 12 
			& 6022.5 & 6025 & 14 & 23 
			& 6022 & 6025.5 & 25 & 30 
			& 6021.5 & 6026 & 34 & 43 \\
13:59:32:397 & 6024 & 6024 & 1 & 2 
			& 6023 & 6024.5 & 14 & 12 
			& 6022.5 & 6025 & 14 & 23 
			& 6022 & 6025.5 & 25 & 30 
			& 6021.5 & 6026 & 34 & 21 \\
13:59:32:413 & 0 & 0 & 0 & 0 
			& 6023.5 & 6024.5 & 8 & 12 
			& 6023 & 6025 & 15 & 23 
			& 6022.5 & 6025.5 & 15 & 30 
			& 6021.5 & 6026 & 34 & 21 \\
13:59:32:430 & 6024 & 6024.5 & 1 & 14 
			& 6023.5 & 6025 & 8 & 23 
			& 6023 & 6025.5 & 15 & 30 
			& 6022.5 & 6026 & 15 & 21 
			& 6022 & 6026 & 25 & 21 \\
13:59:32:443 & 6024 & 6024.5 & 1 & 14 
			& 6023.5 & 6025 & 8 & 23 
			& 6023 & 6025.5 & 15 & 30 
			& 6022.5 & 6026 & 16 & 21 
			& 6022 & 6026.5 & 25 & 15 \\
13:59:32:460 & 6024 & 6024.5 & 1 & 13 
			& 6023.5 & 6025 & 8 & 22 
			& 6023 & 6025.5 & 15 & 30 
			& 6022.5 & 6026 & 16 & 22 
			& 6022 & 6026.5 & 26 & 16 \\
13:59:32:477 & 6024 & 6024.5 & 2 & 13 
			& 6023.5 & 6025 & 10 & 22 
			& 6023 & 6025.5 & 15 & 30 
			& 6022.5 & 6026 & 16 & 22 
			& 6022 & 6026.5 & 32 & 16 \\
13:59:32:490 & 6024 & 6024.5 & 3 & 11 
			& 6023.5 & 6025 & 10 & 22 
			& 6023 & 6025.5 & 15 & 28 
			& 6022.5 & 6026 & 16 & 17 & 6022 & 6026.5 & 32 & 17 \\
\hline
\hline
\end{tabular}
\end{scriptsize}

\end{center}
\caption{Consecutive LOB state snapshots, based on the example of Figure \ref{fig:raw_LOB}.  
FDAX 20 August, 2010. T is the time-stamp, BP and AP are respectively the bid and ask 
limit order prices, BS and AS are respectively the bid and ask sizes at the given 
level.}
\label{fig:formated_LOB}
\end{figure}

This is an example of how LOB is organized for further use. It is based on data from 
Figure \ref{fig:raw_LOB} and shows the state of the LOB at each point in time,  
whenever the LOB has been updated. It is assumed that the LOB was blank before 
the first time-stamp from the excerpt in Figure \ref{fig:raw_LOB}. What follows
is simply a horizontal collection of all orders with the same time-stamp. If
nothing has changed at a given level, then simply the state from the previous 
time step is copied. A cancellation would not be explicitly given, but would 
result in orders from higher levels in the LOB taking up the place of the cancelled order. If
there are no orders at higher levels, then the cancelled order would be filled with zeros.
 Notice that several updates per time step are possible,
including changes of the level of existing orders, incoming new orders, cancellation of
previous orders, changes in the price or the order size of orders at a given level, 
and also execution of orders by matching, or submission of a market order by 
a new market participant. Observe the order execution at 13:59:32:397 and how in the next 
period the exchange does not quote a next best offer. Also look at how the orders from 
upper ask levels assume lower levels immediately after the execution of what appears to 
have been a market order at 13:59:32:397, as now the previous best bid offer of 6023.5
reappears at level one, instead of level zero as the second best bid offer.
\end{landscape}

\newpage

\bibliographystyle{amsalpha}
\bibliography{myreferences}		
\end{document}